\documentclass[a4paper,longbibliography,english,reprint,aps,superscriptaddress,nofootinbib]{revtex4-2}

\usepackage{babel,calc,amsmath,amsthm,amssymb,bm,graphicx,siunitx}
\graphicspath{{images/}{images/SM/}}

\usepackage[T1]{fontenc}
\usepackage{mathdots}
\usepackage{geometry}
\usepackage[unicode=true]{hyperref}
\usepackage{booktabs}
\usepackage{mathtools}
\renewcommand{\arraystretch}{1.2}
\usepackage{multirow}
\usepackage{lipsum}
\usepackage[normalem]{ulem}
\usepackage[dvipsnames]{xcolor}

\usepackage[ruled]{algorithm2e}
\SetKwInput{KwData}{Input}
\SetKwInput{KwResult}{Output}
\SetKwComment{Comment}{~\#~}{}

\hypersetup{
     colorlinks=true,               
     linkcolor=Blue,            
     citecolor=Blue,             
     urlcolor=Blue,          
 }

\newcommand{\braket}[1]{\left\langle#1\right\rangle}

\DeclareMathOperator{\Tr}{Tr}
\renewcommand{\Re}{{\rm Re}}
\renewcommand{\Im}{{\rm Im}}

\newtheorem*{corollary*}{Corollary}

\newtheorem*{proposition*}{Proposition}
\newtheorem{theorem}{Theorem}
\newtheorem*{theorem*}{Theorem}

\theoremstyle{definition}
\newtheorem{definition}{Definition}
\newtheorem*{definition*}{Definition}

\theoremstyle{remark}

\newtheorem*{lemma*}{Lemma}

\newtheorem*{remark*}{Remark}

\newtheorem*{example*}{Example}

\newif\ifdebug

\debugtrue

\ifdebug

\newcommand\delete{\bgroup\markoverwith{\textcolor{Maroon}{\rule[0.5ex]{2pt}{0.4pt}}}\ULon}

\else

\newcommand{\note}[1]{\ignorespaces}
\newcommand{\delete}[1]{\ignorespaces}

\fi

\begin{document}
\renewcommand{\figurename}{Fig.}

\newcommand{\extfig}{Extended Data Figure}
\newcommand{\methodsname}{Methods}
\newcommand{\smname}{Supplementary Material}

\renewcommand{\sectionautorefname}{Sec.}
\renewcommand{\tableautorefname}{Table}
\renewcommand{\equationautorefname}{Eq.}


\title{Quantum learning advantage on a scalable photonic platform}

\affiliation{Center for Macroscopic Quantum States (bigQ), Department of Physics, Technical University of Denmark, Fysikvej, 2800 Kongens Lyngby, Denmark}

\author{Zheng-Hao~Liu$^{\,\P}$}
\let\oldfootnote\thefootnote
\renewcommand{\thefootnote}{$\P$}
\footnotetext{\href{mailto:zheli@dtu.dk}{zheli@dtu.dk}}
\let\thefootnote\oldfootnote
\thanks{These authors contributed equally to this work.}
\author{Romain~Brunel}
\thanks{These authors contributed equally to this work.}
\author{Emil~E.~B.~{\O}stergaard}
\author{Oscar~Cordero}
\affiliation{Center for Macroscopic Quantum States (bigQ), Department of Physics, Technical University of Denmark, Fysikvej, 2800 Kongens Lyngby, Denmark}

\author{Senrui~Chen}
\author{Yat~Wong}
\affiliation{Pritzker School of Molecular Engineering, The University of Chicago, Chicago, Illinois 60637, USA}

\author{Jens~A.~H.~Nielsen}
\author{Axel~B.~Bregnsbo}
\affiliation{Center for Macroscopic Quantum States (bigQ), Department of Physics, Technical University of Denmark, Fysikvej, 2800 Kongens Lyngby, Denmark}

\author{Sisi~Zhou}
\affiliation{Perimeter Institute for Theoretical Physics, Waterloo, Ontario N2L 2Y5, Canada}
\affiliation{Department of Physics and Astronomy and Institute for Quantum Computing, University of Waterloo, Ontario N2L 2Y5, Canada}

\author{Hsin-Yuan~Huang}
\affiliation{Google Quantum AI, Venice, CA, USA}
\affiliation{Institute for Quantum Information and Matter, California Institute of Technology, Pasadena, CA 91125, USA}
\affiliation{Center for Theoretical Physics, Massachusetts Institute of Technology, Cambridge, MA 02139, USA}

\author{Changhun~Oh}
\affiliation{Department of Physics, Korea Advanced Institute of Science and Technology, Daejeon 34141, Korea}

\author{Liang~Jiang}
\affiliation{Pritzker School of Molecular Engineering, The University of Chicago, Chicago, Illinois 60637, USA}

\author{John~Preskill}
\affiliation{Institute for Quantum Information and Matter, California Institute of Technology, Pasadena, CA 91125, USA}

\author{Jonas~S.~Neergaard-Nielsen}
\author{Ulrik~L.~Andersen}
\email{ulrik.andersen@fysik.dtu.dk}
\affiliation{Center for Macroscopic Quantum States (bigQ), Department of Physics, Technical University of Denmark, Fysikvej, 2800 Kongens Lyngby, Denmark}

\date{\today}

\begin{abstract}
    Recent advancements in quantum technologies have opened new horizons for exploring the physical world in ways once deemed impossible. Central to these breakthroughs is the concept of quantum advantage, where quantum systems outperform their classical counterparts in solving specific tasks. While much attention has been devoted to computational speedups, quantum advantage in learning physical systems remains a largely untapped frontier. Here, we present a photonic implementation of a quantum-enhanced protocol for learning the probability distribution of a multimode bosonic displacement process. 
    By harnessing the unique properties of continuous-variable quantum entanglement, we obtain a massive advantage in sample complexity with respect to conventional methods without entangled resources.
    With approximately $5$\,dB of two-mode squeezing---corresponding to imperfect Einstein--Podolsky--Rosen (EPR) entanglement---we
    learn a 100-mode bosonic displacement process using 11.8 orders of magnitude fewer samples than a conventional scheme.
    Our results demonstrate that even with non-ideal, noisy entanglement, a significant quantum advantage can be realized in continuous-variable quantum systems. This marks an important step towards practical quantum-enhanced learning protocols with implications for quantum metrology, certification, and machine learning.
\end{abstract}

\maketitle


Learning the properties of a physical system by performing measurements on it is at the foundation of natural sciences.
In conventional settings, this typically involves collecting a large set of independent measurements of certain variables of the system and applying statistical methods on a classical computer to estimate their underlying distribution, from which the properties of the system can be inferred. However, in quantum systems, the learning task is hindered by the constraints of quantum physics, such as the inherent quantum noise associated with measurements, encapsulated by Heisenberg's uncertainty principle. Consequently, the sample complexity---the number of experiments required to learn certain properties of quantum systems---can scale exponentially with the system size, rendering some learning tasks practically infeasible using classical, conventional learning approaches~\cite{Banchi2023,Anshu2023}.

As an alternative to the conventional approach of using independent probe states and a classical processor for data analysis, quantum learning strategies have been proposed~\cite{Huang2021,Aharonov2022,Chen2022Pauli,Caro2022,Rossi2022,Oh2024,Chen2022,seif2024entanglement}. In such approaches, the probe states are not measured independently but instead undergo a collective quantum algorithmic measurement before data analysis is conducted. By leveraging quantum coherence of the probe states and collective measurements, it has been shown that, for certain finite-dimensional quantum systems, the sample complexity required to learn some of their underlying properties can be dramatically reduced, with an exponentially smaller number of experiments required to complete the task~\cite{Huang2021,Chen2022}. Building on these ideas, quantum advantage in learning was first demonstrated on a superconducting electronic 
platform~\cite{Huang2022}. By utilizing 20 qubits as a probe state, a learning task was accomplished using approximately $10^5$ fewer samples than conventional methods. A more recent experimental exploration of quantum learning advantage with superconducting qubits is given in Ref.~\cite{seif2024entanglement}. Given the system size limit for superconducting systems and the challenges in capturing and transducing unknown quantum states into superconducting qubits, it is particularly intriguing to address how a scalable quantum learning advantage can be achieved in more practical scenarios.

\begin{figure*}
    \centering
    \includegraphics[width=1.5\columnwidth]{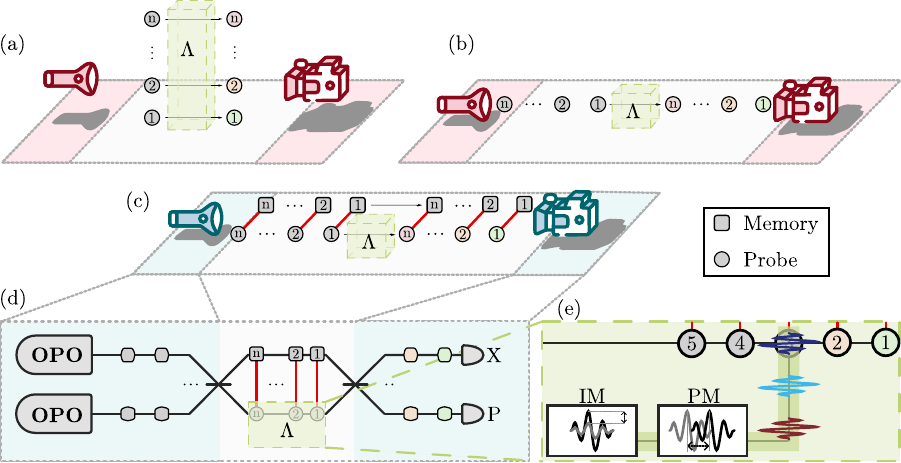}
    \caption{\textbf{Quantum entanglement-enhanced learning with photons.} (a) Conventional learning of a channel. A multi-mode probe state is sent through a channel, followed by a measurement of the probe state to extract the information about the channel. (b) Conventional learning of a multi-time physical process, where the measurement settings are allowed to be adaptive within a sample. We show that for both (a) and (b), a fundamental entanglement-free complexity bound applies to the required sample overhead for the learning task. (c) Quantum entanglement-enhanced learning of a multi-time process. The probe state is allowed to be entangled with an external memory state. The joint measurement of both states makes overcoming the classical complexity limit possible. (d) Implementation of quantum learning with squeezed light. Two-mode squeezing is generated by interfering the outputs of two optical parametric oscillators (OPOs). One of the spatial modes is temporally-multiplexed and used as the probe state while the other is used as the memory. The physical process to be learned is a phase-space random displacement. A Bell measurement between the corresponding modes of the probe and memory states works to extract the information. (e) We realize the displacement process by mixing a frequency-shifted coherent state into the probe. The frequency-shifted coherent state is shaped by two electro-optic modulators, an intensity modulator (IM) and a phase modulator (PM).}
    \label{fig:concept}
\end{figure*}

In our work, we significantly advance the frontier of quantum learning by demonstrating an unprecedented quantum advantage, achieving orders-of-magnitude improvement on a scalable continuous-variable (CV) photonic platform. Using an ensemble of imperfect EPR entangled states of light and a joint CV measurement approach, we successfully learn the amplitude and phase distributions of a multi-mode displacement process with $10^{11}$ times fewer samples than required by an approach without entanglement. Furthermore, our photonic platform enables a significant quantum advantage in distinguishing two families of quantum processes.
Our implementation, capable of learning an infinite-dimensional joint displacement process spanning over 100 modes, tackles problems whose complexity substantially surpasses the previous superconducting qubit-based demonstration~\cite{Huang2022}.
The photonic CV platform~\cite{Weedbrook2012}, a natural architecture for quantum information processing, has been at the forefront in advancing quantum technologies---from boson sampling~\cite{Zhong2020,Madsen2022} to quantum communication~\cite{Takeoka2014,Pirandola2017,Pirandola2020}, computation~\cite{Larsen2021,Tzitrin2021,Larsen2021PRX}, and sensing~\cite{Tse2019,Guo2020,Nielsen2023}. This work demonstrates how photonic systems,  with their established capabilities in computation and sensing, can be leveraged to enhance our ability to learn about physical systems. Moreover, while photonic systems have previously demonstrated their potential in various areas of quantum information processing, a definitive quantum advantage in such systems has not been achieved until now~\cite{Oh2024gbs}. Our achievement thus represents an important milestone in both quantum learning and the broader field of quantum information science.

\textit{Context.}---The task of quantum learning proceeds as follows: The experimenter aims to learn a specific property of a quantum system, such as the probability distribution of a quantum state or the noise characteristics of a particular quantum device. The experimenter probes the device $N$ times, yielding $N$ data samples (See Fig.~\ref{fig:concept}(a)), from which the target property or probability distribution is reconstructed with a certain precision or classified with a specified confidence.

In this work, we focus on the task of learning the properties of random $n$-mode phase-space displacement processes, which model the physical process of random amplitude and phase noise in bosonic channels. These channels are of particular interest because any CV noise channel can be tailored into a random displacement channel by twirling with displacement operators, similar to Pauli twirling in DV systems~\cite{wallman2016noise}. Moreover, learning the properties of multi-time displacement processes has broad applications including gravitational wave detection~\cite{Ganapathy2023,valahu2024}, Raman spectroscopy~\cite{Cheng2015,deAndrade2020}, dark matter searches~\cite{Backes2021,Brady2022}, and microscopic force sensing~\cite{Barzanjeh2022}. 

We learn the dynamical displacement process, labeled $\Lambda$, by probing it with quantum states of light, followed by measurements that extract information about the probability distribution $p(\alpha)$ of the $n$-mode displacement process, where $\alpha$ is the $n$-dimensional complex-valued vector describing the phase-space displacement. We refer the readers to the \smname\ Sec.\,IIA for the formal definition of the displacement process. The key challenge is to resolve sub-Planck features of phase space with the highest possible resolution. This is accomplished through a sequence of $N$ samples using carefully chosen probe states and measurement schemes to reconstruct $p(\alpha)$ or test properties of the displacement process.



The fundamental challenge in this learning task can be understood through the characteristic function $\lambda(\beta)$, which is defined over the dual space of the $n$-mode displacement channel. Mathematically, $\lambda(\beta)$ is the Fourier transform of the phase-space probability distribution $p(\alpha)$. The high-frequency components of $\lambda(\beta)$, i.e., regions where $|\beta|^2$ is large, encode the fine structure of the displacement channel. In a previous work by some of us~\cite{Oh2024}, we proved that learning $\lambda(\beta)$ to a fixed accuracy within a hyperball of squared radius $|\beta|^2 \propto n$ requires a number of samples scaling exponentially in the number of modes $n$. Here, we extend this result to prove (cf. \smname, Theorem 3) that the same lower bound for sample complexity, hereafter referred to as the classical complexity bound, still applies even when the channel is implemented as a multi-time process (as depicted in Fig.~\ref{fig:concept}(b)), and an adaptive strategy is allowed for the measurement. Notably, this exponential scaling in sample complexity is fundamental---it holds for any choice of probe state that is not entangled with an external quantum memory.

\textit{Quantum-enhanced learning.}---We use entanglement to overcome this limitation on learning a random displacement process. Our quantum-enhanced learning scheme, illustrated in Fig.\,\ref{fig:concept}(c), fundamentally differs from conventional entanglement-free approaches: each probe mode in the probing state is entangled with a corresponding auxiliary memory mode, forming EPR entangled (or two-mode squeezed) states of a certain squeezing level. 
The learning process is carried out by sending the probe modes through the displacement process, and subsequently performing pairwise CV Bell measurements on the probe and memory modes. These Bell measurements are joint measurements that simultaneously reveal the quantum correlations between the two modes' amplitude and phase quadratures, thereby bypassing the limitation imposed by the uncertainty principle for individual measurements. 
In \smname\ Sec.\,IIB, we describe the method for estimating the characteristic function from the measured samples.
It allows for efficient extraction of information about the displacement process with a phase-space resolution given by the amount of entanglement. For instance, in the ideal case where the probe and memory modes form perfect EPR states (corresponding to infinite squeezing), the quadratures of the probe and memory modes before the random displacement are perfectly correlated. Any observed deviation from these perfect correlations during the Bell measurement can be attributed solely to the effect of the displacement process. In this scenario, the protocol can extract the complete information of the process without introducing additional noise, demonstrating its potential for noise-free learning.

We realize the quantum-enhanced learning protocol in a CV optical setup, as illustrated in Fig.\,\ref{fig:concept}(d), with a detailed description provided in \smname\ Sec.\,I. The two-mode squeezed vacuum states, comprising the probe and memory modes, are generated by interfering the outputs of two optical parametric oscillators (OPOs). 
Each mode is defined in the temporal domain of the squeezed light, represented by its electromagnetic field weighted by a mode function---a Gaussian-modulated  (full-width half-max of 320 ns) 3.8\,MHz sine wave. Multiple modes are created consecutively at the continuously generated OPO outputs, enabling us to scale up the number of modes for the learning task and achieve a substantial quantum advantage. 
The displacement process is implemented by mixing a weak coherent state into the probe modes via an unbalanced beam splitter (cf. Fig.\,\ref{fig:concept}(e)). By matching the coherent state's temporal mode function and varying its complex amplitude using a pair of electro-optic modulators, we induce a multi-mode correlated displacement. 
To extract the displacement information, we perform Bell measurements by interfering the probe and memory modes and using homodyne detection to measure the amplitude and phase quadratures of the resulting output signals. Our OPOs achieve up to 68\% reduction of noise power during Bell measurements, enabling high precision in detecting the effects of the process.

\begin{figure*}[t]
    \centering
    \includegraphics[width=\textwidth]{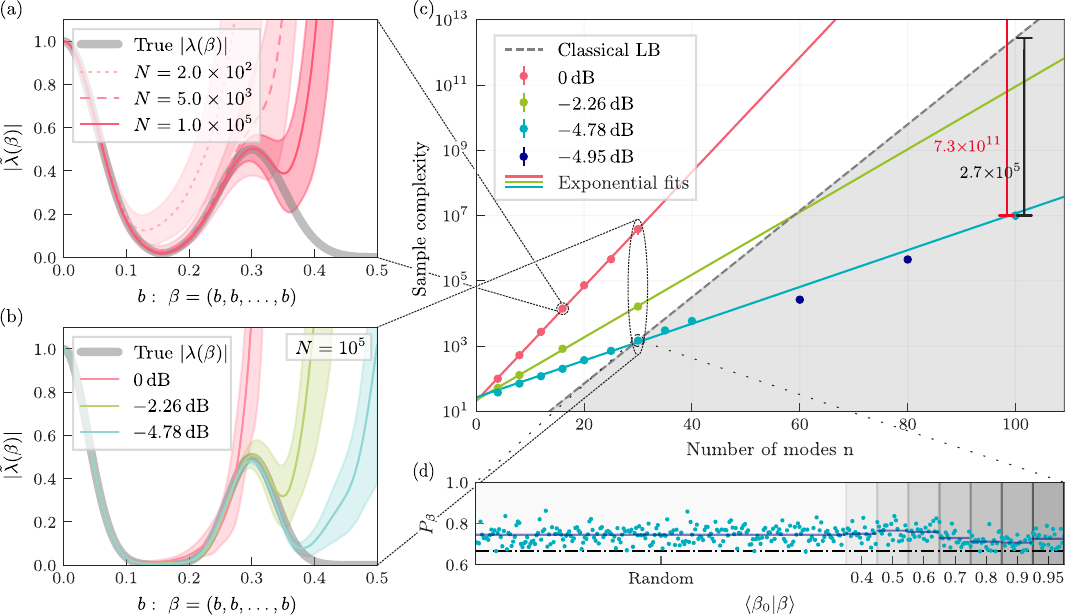}
    \caption{\textbf{Reconstruction of a physical process.} 
    (a) Experimentally reconstructed characteristic function $\tilde{\lambda}({\beta})$ of an $n=16$-mode three-peak process (defined in \smname, Definition S4) with fixed parameters using entanglement-free strategies, compared with the true characteristic function $\lambda({\beta})$. 
    The lines (shadings) show the average outcome ($1\sigma$ standard deviation) of 100 runs of the reconstruction task using different numbers of samples.
    (b) Same as above, but using entangled probe states. Here the number of modes is $n=30$ and we always use $10^5$ samples for the same task.
    (c) Required number of samples to $\epsilon$-close reconstruct $\lambda({\beta})$ of the three-peak process along the $\beta_0$ direction with a success probability of $1-\delta = 2/3$, versus the number of modes. The points are determined from experimental results, and the $1\sigma$ standard deviation error bars are smaller than the data points. The solid lines are log-linear fits. 
    The gray dashed line is the sample-complexity lower bound that applies to any entanglement-free strategy that can learn all processes in a large family which includes the process we studied. 
    (d) Probability of achieving an $\epsilon$-close reconstruction of the $\SI{-4.78}{dB}$, 30-mode characteristic function for various directions in the dual space. The shading highlights the proximity to the displacement direction $\beta_0$. Each probability is computed using $N=1472$ samples---same as required for an $\epsilon$-close reconstruction in (c). The dashed line indicates the target probability of $1-\delta$. }
    \label{fig:chn-recon}
\end{figure*}

\textit{Process reconstruction.}---We demonstrate the quantum enhancement of the learning task by reconstructing the characteristic function $\lambda(\beta)$ of a class of three-peak displacement processes, defined in \smname, Definition S4, using Bell measurement outcomes. We denote the reconstructed characteristic function as $\tilde{\lambda}(\beta)$. For a fixed squeezing parameter $r$, achieving a given reconstruction accuracy requires $N \sim \exp(2e^{-2r}|\beta|^2)$ samples, which grows exponentially with $|\beta|^2$. When the number of samples $N$ is insufficient, the reconstructed characteristic function $\tilde{\lambda}(\beta)$ can diverge at large $|\beta|^2$. To visualize this divergence behavior, we plot the reconstructed characteristic function along a slice $\beta_0 = b\,(1,1,\ldots,1), b\in[0, 0.5]$ in the high-dimensional dual space in Fig.\,\ref{fig:chn-recon}(a): without squeezing ($r=0$), the reconstructed 16-mode characteristic function from 200 samples diverges rapidly and fails to capture the ground truth's peak at $b=0.3$. Furthermore, extending the radius $|\beta|$ of the reconstructable hyperball requires a substantial increase in sample complexity: approximately $10^5$ samples are needed to reveal the peak at $b=0.3$.


The introduction of moderate entanglement through two-mode squeezing ($r > 0$) dramatically reduces this sample complexity. As shown in Fig.\,\ref{fig:chn-recon}(b), when the number of modes increases to 30, reconstruction without entanglement ($r = 0$) fails even with $10^5$ samples, requiring an impractically large number of samples. However, with moderate entanglement, faithful reconstruction becomes achievable with the same number of samples.
To systematically investigate the effect of entanglement on sample complexity across different numbers of modes, we measure the two-mode squeezed vacuum state under various $n$-mode displacement processes using three distinct two-mode squeezing levels: $\SI{0}{dB}$, $\SI{-2.26}{dB}$, and $\SI{-4.78}{dB}$. Notably, during the 4.78\,dB realization, the squeezing increased to 4.95\,dB for mode numbers 60 and 80. 
We characterize the reconstruction performance using $(\epsilon, \delta)$-complexity: the number of samples required to reconstruct $\lambda(\beta)$ with precision $\epsilon$, such that $|\tilde{\lambda}(\beta)-\lambda(\beta)|<\epsilon$ for all $|\beta|\leq0.3\sqrt{n}$, with success probability $1-\delta$. We determine the sample complexity through Monte Carlo resampling of the measurement results. Further details of the process reconstruction experiment are provided in \smname\ Sec.\,IV.

As shown in Fig.\,\ref{fig:chn-recon}(c), entanglement in the form of two-mode squeezing significantly enhances the scaling of sample complexity, leading to substantial improvements as the number of modes increases. 
For $n=100$, the sample complexity with the strongest squeezing ($\SI{-4.78}{dB}$) remains about $10^7$. In contrast, from an exponential fit of low mode number data, we estimate that the entanglement-free scheme using vacuum states and heterodyne measurements would require a sample complexity as high as $7.3\times10^{18}$ for the same task, or more than twenty million years for acquiring all the data if the samples are generated at the same rate (1\,MHz per mode) as in our experiment. This represents an empirical improvement of 11.8 orders of magnitude. 

Further, in Fig.\,\ref{fig:chn-recon}(d) we confirm that the success probability of reconstructing $\lambda(\beta)$ is the lowest along directions $\beta$ that are close to $\beta_0$, i.e. in the direction of the distribution's high-frequency peaks.
Therefore, the sample complexity in Fig.\,\ref{fig:chn-recon}(c) is a faithful estimation of the true complexity for learning this channel. 


\begin{figure*}[ht]
    \centering
    \includegraphics[width=\textwidth]{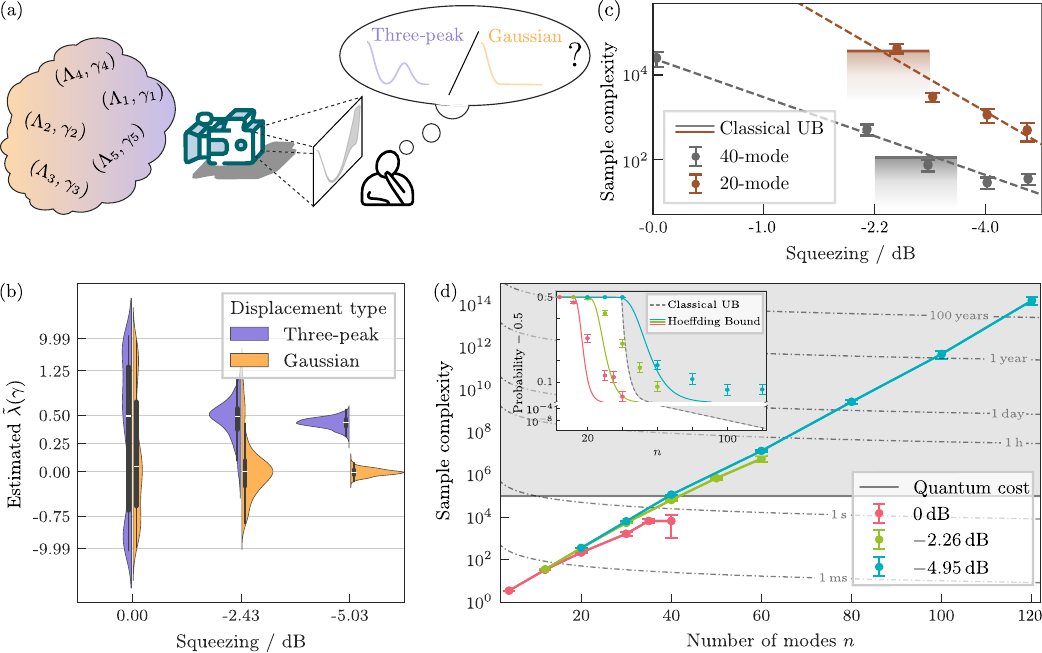}
    \caption{\textbf{Hypothesis testing.} (a) The objective is to distinguish whether a displacement process belongs to the three-peak family with an unknown parameter or the Gaussian family. (b) An example of the separation of the estimator, $\tilde{\lambda}(\beta=\gamma)$, for two types of 40-mode displacement processes using different amounts of squeezing. In the noiseless case, the value is expected to be 0.5 (0) for three-peak (Gaussian) channel. (c) Sample complexity for achieving $2/3$ success probability in a $\kappa=0.2$ hypothesis test, estimated with varying amounts of squeezing. The solid and dashed lines indicate the classical complexity bound for achieving the same success probability and the exponential fit, respecively. The shading indicates the region excluded by the classical complexity bound (see \smname Sec.\,VB for more details). (d) Inset: the measured probability of winning the $\kappa=0.2$ hypothesis testing game for different numbers of modes, using $10^5$ samples and various amounts of squeezing. Solid lines represent a pessimistic estimation of success probability derived from the Hoeffding bound\,\cite{Oh2024}.
    Main: Minimum sample complexity for any conventional strategy to achieve the same success probability as reported in the inset, calculated according to the classical complexity bound, and the corresponding sample collection time at a 1\,MHz/mode rate.
    Error bars represent the $1\sigma$ standard deviation from a 25-step sequential sampling. 
    The shaded region indicates the existence of a quantum advantage.}
    \label{fig:hypo-test}
\end{figure*}

\textit{Provable, scalable quantum advantage.}---Also shown in Fig.\,\ref{fig:chn-recon}(c) is a lower bound on the sample complexity of entanglement-free learning schemes, derived in \smname\ Sec.\,IIC. Compared to this bound, the entanglement-enhanced reconstruction of the 100-mode process uses more than five orders of magnitude fewer samples. However, the sample-complexity lower bound applies to entanglement-free schemes that can learn the characteristic function $\lambda(\beta)$ accurately for any random displacement process in a large family, and for all values of $\beta$ in a specified bounded range, while our experiment learns the characteristic function for processes chosen from a smaller family and for more restricted values of $\beta$. Therefore, the improved process reconstruction we achieved using entanglement does not demonstrate a provable quantum advantage. 

To establish a provable quantum advantage, we consider the task of identifying specific features of an unknown process. 
Specifically, we design a hypothesis testing game (cf.\ Fig.\,\ref{fig:hypo-test}(a)) where a dealer prepares several $n$-mode displacement processes $\Lambda_k, k\in\{1, \ldots, K\}$. Each process may or may not exhibit a feature---specifically, two peaks at the locations $\pm\gamma_k$ in its characteristic function. Depending on the existence or absence of these peaks, the processes are grouped into three-peak and Gaussian families, respectively (cf.\ \smname, Definitions 4 and 5). A challenger is allowed to implement each process $N$ times. After all the measurements are finished, the dealer reveals $\gamma_k$ and the challenger is asked to identify the family that the process belongs to.
In the quantum strategy for the game, the challenger records the Bell measurement outcomes and computes the estimator $\tilde{\lambda}(\beta)$ at $\beta=\gamma_k$ once the value is announced. They then compare the value of $\tilde{\lambda}(\beta)$ with a threshold $\lambda_0$ to classify the process. If $\tilde{\lambda}(\beta)>\lambda_0$ is observed, they will guess the process as three-peak type, and otherwise as Gaussian.
The random nature of $\gamma_k$ in the hypothesis testing game removes the excessive prior information. Consequently, an experimentally measured sample complexity surpassing the classical complexity bound can demonstrate a provable quantum advantage.
On the other hand, the quantum strategy remains effective even if the challenger is unaware of $\gamma_k$ during the measurement process, highlighting the practicality of the approach.  

We conducted a series of hypothesis testing games to conclusively observe the provable quantum advantage, with full details provided in \smname\ Sec.\,4. In these games, the average distance of the peaks from the origin is proportional to $\sqrt{\kappa n}$, where the resolution constant $\kappa=0.2$ controls the hardness of the task. 
First, we fixed the number of samples at $N=10^5$ and classified a set of 40-mode processes using different levels of two-mode squeezing. The behavior of the estimators across these experiments is reported in Fig.\,\ref{fig:hypo-test}(b). As the squeezing increases, the displacement signal becomes more pronounced against the noise, and the estimator distribution evolves from being almost random to strongly clustering around the true value. 
Next, we used the Monte Carlo method to determine the number of samples $N$ required to achieve a success probability of $2/3$ in 20-mode and 40-mode hypothesis testing games. As shown in Fig.\,\ref{fig:hypo-test}(c), squeezing reduces the sample complexity in these tasks. Even at these modest mode numbers, a quantum advantage is clearly observed.

Finally, we explore how the degree of quantum advantage scales with the number of modes of the displacement process. To do this, we measured the success probability of hypothesis testing with a fixed sample size of $N=10^5$, then calculated the number of classical samples required to achieve the same success probability. A higher success probability corresponds to a larger equivalent classical sample complexity and demonstrates a stronger quantum advantage.
As the number of modes increases, process classification becomes more challenging because the estimator diverges more rapidly for even smaller values of $\beta$. This behavior is illustrated in  
Fig.\,\ref{fig:hypo-test}(d) inset, which shows that the success probability evolves through three phases for different squeezing levels: starting from $\simeq 1$ in low mode number systems where the information from the displacement process dominates the noise,
transitioning through a region of decline, and eventually converging to $\simeq 0.5$ when noise prevails and the estimator ``diverges'' at $\beta = \gamma$, making the estimate no better than a coin toss. 
Importantly, increased squeezing shifts the transition region to higher mode numbers, allowing some data points to achieve success probability significantly above what is possible with classical strategies using the same number of samples.
When the success probability significantly exceeds $0.5$, we compute the equivalent classical sample complexity and compare with the $10^5$ realized samples, which took $n \times$ 0.1\,s to acquire. Our results, shown in Fig.\,\ref{fig:hypo-test}(d), demonstrate the exponential scaling of classical sample complexity with the number of modes and emphasize how the quantum advantage increases as the number of modes grows.
For the largest scale (120-mode) experiment, we measured a hypothesis testing success probability of $0.563\pm 0.025$, exceeding the bound for conventional strategy ($0.5+3.8\times10^{-11}$) with a confidence level of $99.3\%$. To learn one process and achieve the same success probability with a conventional approach, $1.6\times10^{14}$ classical samples would be required. This translates into an expected measurement time of more than 600 years at a 1\,MHz rate. Our result thus indicates a provable quantum advantage of $9.2$ orders of magnitude.

\textit{Outlook.}---In this work, we have demonstrated a substantial quantum advantage in learning using a scalable, albeit noisy and lossy, photonic platform based on EPR entangled states and CV Bell measurements. Despite system losses of approximately 20\%, we achieved a quantum improvement exceeding 11 orders of magnitude compared to the classical approach of using coherent state probes and heterodyne detection. This translates into an expected reduction of measurement time from more than twenty million years to less than 15 minutes, assuming the modes are generated at 1\,MHz frequency as in our experiment. Our method therefore enables the resolution of intricate features in highly complex systems, offering to potentially uncover hidden structures that would remain completely inaccessible with classical measurement techniques. Further, our research sheds light on a novel quantum learning framework, where the information to be learned is encoded into the temporal domain. This quantum learning framework invites further exploration both in theory and in experiments.

Looking ahead, our scalable photonic platform lays the foundation for tackling even more complex tasks.  Future advancements in reducing losses, increasing squeezing levels, and integrating advanced photonic technologies could further enhance the capabilities of quantum learning protocols. The insights gained from our approach have the potential to drive significant progress in quantum-enhanced sensing, parameter estimation, and even quantum-enhanced machine learning algorithms, where photonic platforms are uniquely suited to tackle high-dimensional and continuous-variable problems. By leveraging quantum learning, these enhanced protocols could unlock unprecedented opportunities for scientific discovery and technological innovation. 

While previous demonstrations of quantum advantage on noisy intermediate-scale photonic platforms have primarily focused on quantum computational tasks like Gaussian boson sampling, our work showcases a fundamentally different achievement: a significant quantum advantage in learning. This task extends beyond computation to encompass the understanding and characterization of physical systems, highlighting the versatility of photonic platforms and their critical role in advancing quantum technologies. 



    







\let\oldaddcontentsline\addcontentsline
\renewcommand{\addcontentsline}[3]{}
\bibliographystyle{rev4-2mod}
\bibliography{eedl_sm.bib}
\let\addcontentsline\oldaddcontentsline


\vspace{10pt} \noindent \textsf{\textbf{Acknowledgements.}}---We thank Huy Quang Nguyen and Benjamin Lundgren Larsen for the help with the experiment.
We gratefully acknowledge support from
the Danish National Research Foundation, Center for Macroscopic Quantum States (bigQ,\ DNRF0142), 
EU project CLUSTEC (grant agreement no. 101080173),
EU ERC project ClusterQ (grant agreement no. 101055224), 
NNF project CBQS, 
Innovation Fund Denmark (PhotoQ project, grant no.\ 1063-00046A),
and the MSCA European Postdoctoral Fellowships (\href{https://doi.org/10.3030/101106833}{GTGBS, project no.\ 101106833}).
J.P. acknowledges support from the U.S. Department of Energy Office of Science, Office of Advanced Scientific Computing Research (DE-NA0003525, DE-SC0020290), the U.S. Department of Energy, Office of Science, National Quantum Information Science Research Centers, Quantum Systems Accelerator, and the National Science Foundation (PHY-1733907). The Institute for Quantum Information and Matter is an NSF Physics Frontiers Center.
L.J. acknowledges support from the ARO(W911NF-23-1-0077), ARO MURI (W911NF-21-1-0325), AFOSR MURI (FA9550-19-1-0399, FA9550-21-1-0209, FA9550-23-1-0338), DARPA (HR0011-24-9-0359, HR0011-24-9-0361), NSF (OMA-1936118, ERC-1941583, OMA-2137642, OSI-2326767, CCF-2312755), NTT Research, and the Packard Foundation (2020-71479).
S.Z. acknowledges funding provided by Perimeter Institute for Theoretical Physics, a research institute supported in part by the Government of Canada through the Department of Innovation, Science and Economic Development Canada and by the Province of Ontario through the Ministry of Colleges and Universities. 
C.O. acknowledges support from Quantum Technology R\&D Leading Program (Quantum Computing) (RS-2024-00431768) through the National Research Foundation of Korea (NRF) funded by the Korean government [Ministry of Science and ICT (MSIT)].
    






\appendix
\onecolumngrid

\setcounter{equation}{0}
\setcounter{figure}{0}
\renewcommand{\theequation}{S\arabic{equation}}
\renewcommand{\thefigure}{S\arabic{figure}}
\renewcommand{\appendixname}{}
\setlength{\tabcolsep}{3pt}
\renewcommand{\arraystretch}{1.3}
\setlength{\baselineskip}{14pt}
\linespread{1.3}

\bigskip\bigskip\bigskip

\begin{center}
{\large \bf Supplementary Material for \\ ``Quantum learning advantage on a scalable photonic platform''}
\end{center}

\tableofcontents

\section{Experiment}

\begin{figure}[t]
    \centering
    \includegraphics[width=\textwidth]{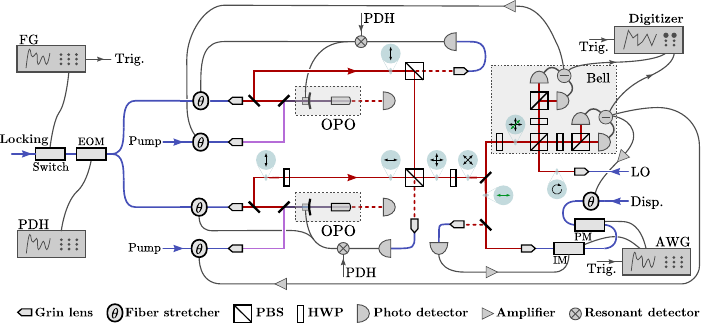}
    \caption{Full experimental setup for quantum-enhanced learning with squeezed light. Two-mode squeezed vacuum (TMSV) is created by combining the outputs of two degenerate OPOs with orthogonal polarizations using a polarization beam splitter (PBS), and then Hadamard-mixing the polarizations with a half-wave plate (HWP). A coherent state at the correct sideband frequency is created by intensity (IM) and phase (PM) modulating a classical laser beam, and the displacement operation is induced by mixing this coherent state on a tapping mirror with one polarization modes of the TMSV. The modulators are driven by an arbitrary waveform generator (AWG) to implement the desired random displacement process. A Bell measurement is performed by reverting the polarization mixing of the two OPOs and interfering them with the two polarization modes of a strong classical beam (local oscillator, LO). A function generator (FG) generates the lock--measure signal, turning off the locking beam during the measurement period and turning it back on during the locking period using a fiber-coupled switch. This signal is also used as a trigger signal for the data acquisition digitizer and the AWG. A second function generator produces the Pound--Drever--Hall sidebands (PDH) on the locking beam by using an electro-optical phase modulator (EOM).}
    \label{fig-sm:setup}
\end{figure}

The setup used to conduct the experiment described in the main text is depicted in Fig.\,\ref{fig-sm:setup}. We generate continuous-wave quadrature-squeezed light at 1550\,nm wavelength with two optical parametric oscillator (OPO) cavities. The design of the OPO is a modified version of the OPO used in\,\cite{Jens19,Jens23} and will be described in \autoref{ssec:opo}. To maintain the correct phase throughout the setup, we use a lock--measure scheme (cf. \autoref{ssec:lock}) to cycle between a locking stage, where a locking beam seeds the OPOs and the feedback controls for phase locks are activated, and the actual measurement stage where the locking beam is turned off and all feedback controls freeze. The cycle occurs at 50\,Hz rate and the measurement runs for 15\,ms (75\% duty cycle).

We couple the output squeezed light from both OPOs into a passively phase-stable free-space polarization interferometer setup using a 99:1 beam splitter, and combine the outputs of the two OPOs on a polarization beam splitter. We have used a reference cavity to carefully align the spatial modes of the two beams.
The two co-propagating single-mode squeezed vacuum (SMSV) beams with orthogonal squeezing quadratures are further rotated from the horizontal- ($H$-) and vertical- ($V$-)polarizations to diagonal- ($D$-) and anti-diagonal- ($A$-)polarizations using a half-wave plate (HWP), forming a two-mode squeezed vacuum (TMSV) on the $H$/$V$-basis. 
We use the $V$-polarized mode as the probe state and the $H$-polarized mode as the memory state in the learning scheme.
Subsequently, we realize the displacement operation by mixing a displacement beam in $V$-polarization into one of the TMSV beams using a 99:1 tapping mirror. We describe the details of the displacement setup in Sec.\,\ref{ssec:disp}. 

To realize the Bell measurement, we rotate the polarization of the beams once again with a HWP at $-22.5^\circ$, so the TMSV returns to two (displaced) SMSVs at $H$- and $V$-polarizations;
The displacement is split evenly between the two polarizations. 
These squeezing beams then mix with a strong coherent state---the local oscillator (LO) with a circular polarization at a PBS, so the $V$-polarized squeezed mode will co-propagate with the $H$-component of the LO and vice versa. Finally, each of the two re-combined modes go through a HWP at 22.5$^\circ$ followed by a PBS, and the two outputs of the PBS are directed to the two ports of a homodyne detector. The polarization setting of the LO causes a $\pi/2$ relative phase between the two squeezed modes with respect to their corresponding LO modes; this way, the two homodyne detectors detect the $x$- and $p$-quadratures of the two squeezed modes, respectively, aligning with their squeezing quadratures.

We use the temporal mode function (cf. \autoref{ssec:modefunc}) to properly define the modes and displacements in a continuous-wave setup. The length of one mode is chosen to be 1\,\textmu s to accommodate the OPO's squeezing bandwidth. The short mode function enables us to multiplex the modes in the time-domain and scale-up the encoded modes without adding to the complexity of the setup: a total of 15,000 modes can be encoded in one lock-measure cycle, and we use 100 modes as the maximum size of a sample that undergoes a displacement process. Finally, in \autoref{ssec:sys-char} we report the characterization result of the setup.

\subsection{Squeezed light source}
\label{ssec:opo}
The squeezed light sources used in this project are two OPOs consisting of a $\SI{10}{\milli \meter}$ periodically poled potassium titanyl phosphate (PPKTP) crystal polished to form a linear hemilithic cavity together with a piezo-actuated meniscus mirror.
The cavity is double resonant for both the $\SI{1550}{\nano \meter}$ squeezing beam and the second-harmonic $\SI{775}{\nano \meter}$ pump beam. The reflectivities on the meniscus mirror for the pump and the locking beam are $73.4\%$ and $94.5\%$ respectively, leading to a finesse of $20$ and $110$.

To achieve the phase lock of the cavity, we need to introduce a locking beam that has the same wavelength as the squeezing beam. Both the pump and the locking beam are sent from the fiber into free-space using a gradient-index (GRIN) lens. They are then combined on a dichroic mirror and coupled into the cavity through the piezo-actuated mirror. The squeezed light is extracted through the same mirror and then we use a 99:1 free-space beam splitter to extract the squeezed mode with high efficiency into the main setup. The parameters of the cavity can be found in \autoref{tab:cavity-parameters}.

\begin{table}[b]
    \begin{tabular}{ccc}
    \toprule\toprule
        Parameters                                   & Pump        & Locking  \\
    \midrule
        Crystal radius of curvature            & \multicolumn{2}{c}{\SI{10}{\milli\meter}}     \\
        Piezo-mirror outer radius of curvature & \multicolumn{2}{c}{\SI{8}{\milli\meter}}      \\
        Piezo-mirror inner radius of curvature & \multicolumn{2}{c}{\SI{18}{\milli\meter}}     \\
        Piezo-mirror diameter                     & \multicolumn{2}{c}{\SI{7.3}{\milli\meter}}    \\
        Reflectivity piezo-mirror              & 73.4\%          & 94.5\%     \\
        Free Spectral Range                    & \SI{4.47}{\giga\hertz}         & \SI{4.51}{\giga\hertz}    \\
        Intra-cavity loss                      & -               & 0.096\%    \\
        Finesse                                & \textless{}20.3 & 109.21     \\
        Half-width-half-maximum                & \SI{101.81}{\mega\hertz}       & \SI{20.35}{\mega\hertz}   \\
        Escape efficiency                      & -             & 98.31\%   \\
    \bottomrule\bottomrule
    \end{tabular}
    \caption{Summary of the cavity component parameters.}
\label{tab:cavity-parameters}
\end{table}

\subsection{Phase locking scheme}
\label{ssec:lock}
To achieve stable phase locking, we operate the system with a lock--measure scheme running at a $\SI{50}{\hertz}$ rate. In the first quarter of the cycle ($\SI{25}{\%}$ duty cycle), a locking beam is turned on to seed the OPO. The output beam propagates through the setup providing phase information for the 
error signals that are actively fed back to piezo actuators in the various phase shifters and cavities of the experiment. For the rest of the cycle, the locking beam is shut down, allowing the cavity to generate a squeezed vacuum, and the feedback signals are held constant. This way, the only source of phase change during the measure period is the noise from the environment. As shown by a sweeping squeezing measurement, the effect of phase drifting during the 15\,ms measurement time is negligible and causes a minimal amount of decrease in the squeezing level. On the other hand, the squeezing level saturates slowly at the beginning of the lock period, due to the finite response speed of the micro electromechanical system (MEMS) optical switch on the locking beam path. Therefore, we only start the measurement 4\,ms after turning off the locking beam.

\begin{figure}
    \centering
    \includegraphics[width=0.8\linewidth]{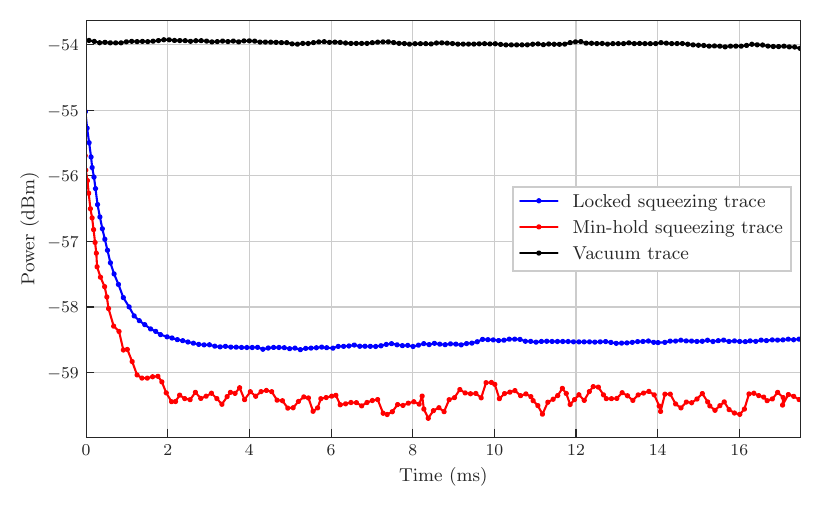}
    \caption{A zero-span noise power measurement of the homodyne detector's output, acquired on a spectrum analyzer centering on the $\SI{3.8}{\mega\hertz}$ frequency sideband.
    The trace is recorded during a sweeping time of $\SI{17.5}{\milli\second}$ starting at the end of the locking period. The min-hold squeezing is from releasing the LO phase lock and capturing the lowest noise power possible. The increase of squeezing at the first $\SI{2}{\milli\second}$ of the trace is due to the response time of the MEMS switch shutting down the locking beam.}
    \label{fig-sm:sq-trace}
\end{figure}

Going through the feedback systems in chronological order, the OPO cavities are locked to resonance using the Pound--Drever--Hall (PDH) technique\,\cite{drever1983laser, black2001introduction}. An electro-optical phase modulator upstream of the cavities creates locking sidebands at $\SI{80}{\mega \Hz}$ for each cavity. The PDH signals are collected by tapping off 1\% of the squeezing beams sampled into homemade fiber-coupled resonant detectors. 
The detector comes with built-in down-mixing functionality and split AC and DC outputs. The output of the AC-port is the input signal down-mixed with $\SI{80}{\mega\hertz}$ sine wave and provides the error signal for the PDH lock, which is sent to a Red Pitaya FPGA board. The built-in lockbox module of PyRPL\,\cite{Neuhaus23} is used as the PID servo to send the feedback to the high-voltage amplifier, which in turn drives the cavity piezo.

To stabilize the relative phase between the pump and the locking beam, we introduce dither signals ($\SI{18}{\kilo \Hz}$ and $\SI{13.3}{\kilo \Hz}$ for the two cavities respectively) to piezo fiber stretchers controlling the locking beam phases before the two OPOs. These sidebands pick up information about the parametric gain, and we extract these by utilizing the $\SI{50}{\kilo\hertz}$ low-pass filtered DC-port of the resonant detector. By down-mixing the DC-port output in an IQ module of PyRPL, we acquire  error signals that are fed back to the same fiber stretchers in the locking beam path. This technique allows us to stabilize the parametric gain of the cavities to the de-amplification point (corresponding to aligning the squeezed quadrature to the amplitude quadrature of the locking beam).  

To lock the relative phase between the LO and the squeezed light, we make use of the fact that the squeezed quadrature is set at the homodyne detector when the de-amplified locking beam is in phase with the LO. Experimentally, we split the output electronic signal of the homodyne detector and down-mix it with the frequency of the parametric gain dithering signals. This allows us to lock one OPO to the de-amplified $x$-quadrature of the locking beam and the other one to the de-amplified $p$-quadrature---the squeezed quadrature of the squeezed vacuum. For these locks, the error signals are fed back to a fiber stretcher in the pump beams.

Finally, the displacement beam is likewise locked to the local oscillator on the homodyne detector by a small dither produced by a fiber stretcher in the displacement beam path of $20$\,kHz, which is down-mixed within PyRPL and fed back to the same fiber stretcher for generating the dither signal.

\subsection{Mode Function}
\label{ssec:modefunc}

\begin{figure}[t]
    \centering
    \includegraphics[width=\linewidth]{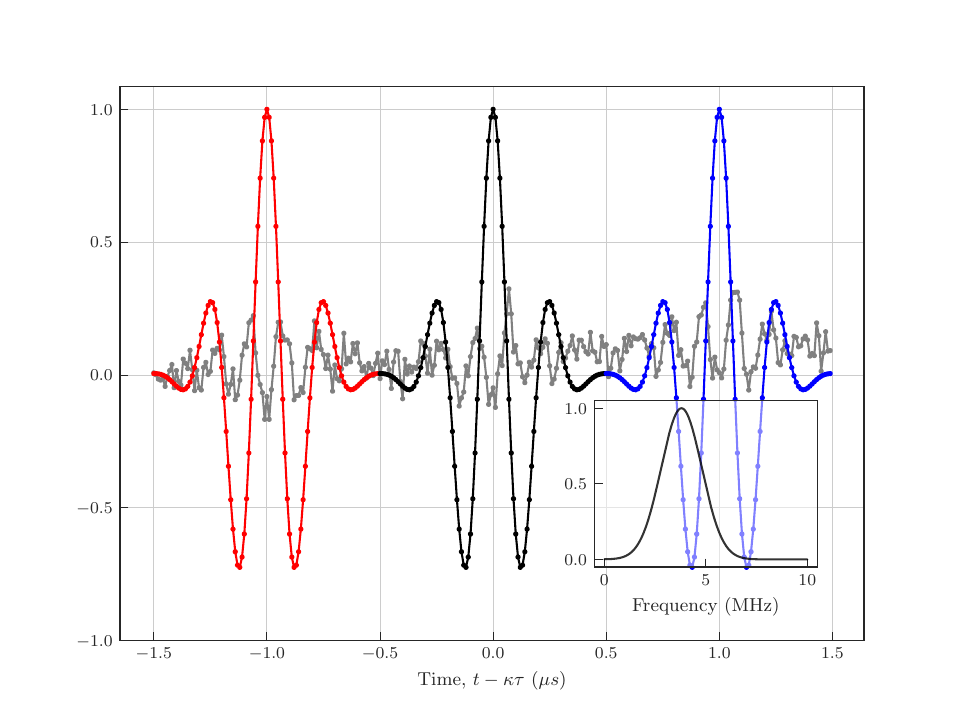}
    \caption{Temporal mode function of three neighboring modes, together with an acquired quadrature time trace (gray). The inset shows the corresponding spectrum of a temporal mode function.}
    \label{fig-sm:mode-func}
\end{figure}

We encode the probe and memory states of the learning scheme in the temporal domain. To map the optical field to the quadrature values, we define the temporal mode function, $f_k(t)$, for the $k$-th individual mode. In the experimental setup, a quadrature $\hat{q}(t)$ is continuously monitored through homodyne detection. By integrating the recorded quadrature time trace, weighted by the temporal mode function, we obtain the measured quadrature for the corresponding temporal mode:
\begin{equation}
    \hat{q}_k = \int f_k(t)\hat{q}(t)\text{d}t
\end{equation}
The temporal mode function defines the temporal mode with a temporal mode duration set to be $\tau = \SI{1}{\micro\second}$. The shape of the mode function is designed to exploit the maximum squeezing and avoid too low frequencies, where the phase noise dominates and the mixing between squeezing and anti-squeezing is more detrimental. In this work, we choose the mode function to be a 3.8\,MHz sine wave, modulated by a Gaussian envelope (full-width half-maximum 320\,ns), and truncated to a length of 1\,\textmu s as the temporal mode function. Explicitly, the mode function is defined as:
\begin{equation}
    f_k(t)= \begin{cases}
    \cos\left(2\pi f_\text{sb}(t-k\tau)\right) e^{-\frac{1}{2}\kappa^2(t-k\tau)^2} , & \text{if } |t-k\tau|<\tau/2 \\
    0 , & \text{otherwise}
    \end{cases}
\end{equation}
where the sideband frequency, $f_\text{sb} = \SI{3.8}{\mega\hertz}$, is optimized to reduce the quadrature variances, and the bandwidth is chosen to be $\kappa = 2\pi\times \SI{1}{\mega\hertz}$. The mode function is a product of a Gaussian function and a sine wave. In the frequency domain, the Gaussian function defines the bandwidth $\kappa$ within the squeezing bandwidth and the sine wave centers the bandwidth around the sideband frequency $f_\text{sb}$. 

\subsection{Displacement setup}
\label{ssec:disp}

The displacement operation is implemented by inserting a displacement beam into the squeezed light through a highly reflective mirror. The displacement beam is a weak coherent state, and its complex amplitude is set by two EOMs: in the $x$-quadrature of the phase space by an IM, and in the $p$-quadrature by a PM. We drive the IM and PM with a waveform that is proportional to the mode function; by adjusting the amplitude of the driving waveform, the desired amount of displacement at the sideband frequency is achieved. 
The intensity of the displacement beam is stabilized by an active control loop, with the DC bias of the IM executing the feedback operation. The displacement beam is mixed into the squeezed mode with a $V$-polarization, controlled by quarter and half wave plates. 

In the experimental setup, the IM and PM are both driven by radio frequency (RF) signals from an arbitrary waveform generator (AWG). Due to electronic delays within the setup, a delay occurs between waveform generation and measurement. To characterize this delay, we send a high-intensity, single-mode displacement to each modulator, and a temporal trace is captured on the homodyne detector. We then perform the convolution of the mode function with the temporal trace; the delay corresponding to the experimental setup is found where this convolution is maximum.

\begin{figure}[tb]
    \centering
    \includegraphics[width=0.8\linewidth]{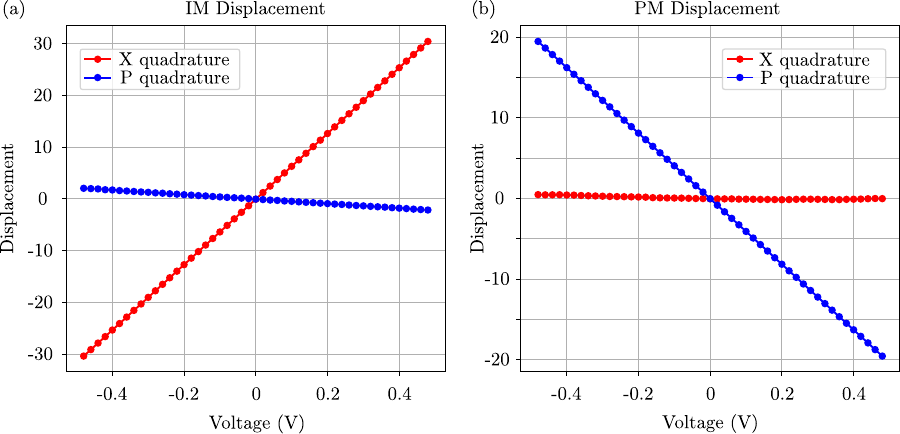}
    \caption{Calibration of the displacement induced by the modulators and measured on both homodyne detectors. The voltage is at the time $t=0$ at the center of the waveform, and each of the displacement values are averaged over $10^4$ runs of the same experiment. (a) Displacement induced by IM measured on both homodynes. The slope on the $p$-quadrature indicates a small crosstalk from the IM into the $p$-quadrature. (b) Displacement induced by PM measured on both homodynes. The slope of the $x$-quadrature indicates negligible crosstalk.}
    \label{fig:disp-cal}
\end{figure}

\textit{Crosstalk mitigation.}---We run calibration tests during the experiment from time to time to determine the effect of displacement induced by the IM and PM. The calibration is made by sending calibration waveforms to the two modulators. The waveform is of the same shape of the mode function in the time-domain. We vary the amplitude of the waveforms and measure the effect of displacement at the Bell measurement site. As can be seen in \autoref{fig:disp-cal}, due to the imperfect manufacturing, the IM also displaces the coherent state in the $p$-quadratures, resulting in a crosstalk that is visible during a Bell measurement. To correct the IM crosstalk, we compensate the effect by subtracting the anticipated amount of crosstalk from the displacement sent by the PM. On the other hand, the cross-talk induced by the PM is negligible. 

\textit{Drift compensation.}---Due to the phase and polarization fluctuations in a real experimental setup, the displacements measured at the homodyne detector may only reflect a distorted version of the process' true distribution, hampering our ability to learn the process. To ensure the precision of the protocol, we encode additional, known displacements and send them together with the train of displacements to be learned. The observed values for these known displacements are used to characterize and correct the imperfection during post-processing.

Our protocol works as follows: for every $10^6$ modes sent to Bob in 100 lock--measure cycles, Alice sends $2000$ modes with high, fixed, and known displacements. Explicitly, half of them are $D_{x_0} = (10, 0)$ along the $x$-quadrature, and the other half are $\Delta_{p_0} = (0, 10)$ along the $p$-quadrature. Suppose, during measurement, the average of the displacements corresponding to those sent as $\Delta_{x_0}$ are $(\Delta_{\{x, x\}}, \Delta_{\{x, p\}})$, and to those sent as $\Delta_{p_0}$ are $(\Delta_{\{p, x\}}, \Delta_{\{p, p\}})$. Due to the linear nature of the noisy displacement process, the effect of the noise must be an affine transformation of these quadrature values, and can be captured by the affine matrix $A$:
\begin{equation}
    A := \frac{1}{10}\begin{pmatrix}
        \Delta_{x,x} & \Delta_{p,x}\\
        \Delta_{x,p} & \Delta_{p,p}
    \end{pmatrix}.
\end{equation}
The affine matrix encodes phase drift and thus the distortion. Given the affine matrix is not too far from identity, we can inverse transform the grid in the characteristic function space and correct the distortion. Explicitly, by applying the correction on the estimator in Eq.\,\eqref{eq:estimator} in the distortion-free case, we found that the process's characteristic function can be determined with the following unbiased estimator:
\begin{equation}
    \Tilde{\lambda}(\beta)=\frac{1}{N}\sum_{i=1}^Ne^{e^{-r_\text{eff}}|A^{-1}\beta|^2}e^{\zeta^{(i)\dagger}(A^{-1}\beta) - (A^{-1}\beta)^{\dagger}\zeta^{(i)}}.
    \label{eq:adapted-estimator}
\end{equation}

Due to the short period of the lock--measure cycle, all the modes in 100 lock--measure cycles go through virtually the same distortion, and we use a single affine matrix to describe the noise behavior. However, after a long run of measurement, the total drift experienced by the setup can become more pronounced, causing an undesirable change in the phase-space grid shape. To prevent this from happening, we use the following termination condition: if the drift is too large $|A|>\SI{10}{\%}$, the acquisition is stopped, and a new round of displacement calibration is taken. 

\begin{figure}[t]
    \centering
    \includegraphics[width=0.9\linewidth]{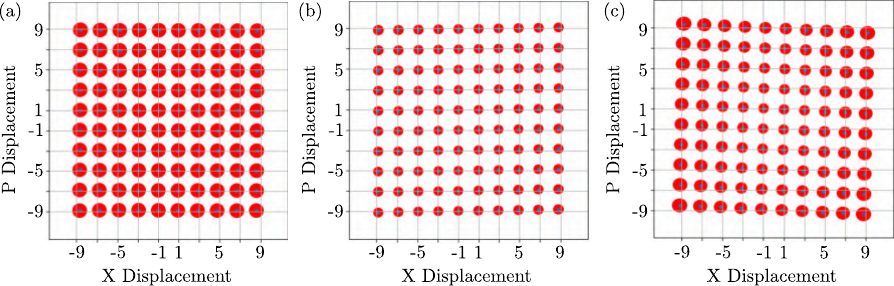}
    \caption{The figures are the result of a diagnosis function, which creates a grid displacement of $-9$ to $9$ in both quadratures. The vacuum variance is set to be $1/2$. A red blob represents the $1\sigma$ covariance matrix of 10000 shots of the same displacement. (a) Measurement of the displacement sent by the diagnosis function with $0\,$dB squeezing, corresponding to an affine matrix close to the identity matrix.  (b) Measured displacement with $4\,\text{dB}$ squeezing, matching the desired displacement and also corresponding to an affine matrix close to the affine matrix as the identity matrix. (c) Results of the diagnosis function with an intentional ground loop in the locking system, defining the affine matrix as a skewed matrix.}
    \label{fig-sm:diag}
\end{figure}

To certify the precision of displacement operation, we also implement a diagnosis test. The test is done by setting the process to implement a set of fixed displacement, $D(x_0+i\,p_0), (x_0,p_0)\in\{-9, -7, -5, \ldots, 7, 9\}$, and extract the quadrature values of the probe states undergone the set of processes using Bell measurement. In \autoref{fig-sm:diag}, we visualize the results of the diagnosis test by plotting the covariance matrices of 10000 shots of the same displacement as ellipsoids. We see that when using TMSV as probe and memory states, the uncertainty of the extracted quadrature values clearly decreases. Finally, when the locking system is not set up properly, the phase diffusion will cause the uncertainty of larger displacements to be higher than small displacements. The diagnosis function thus provides information about how the entire system performs, and we use it to keep track of the experiment over a long period of data collection.

\subsection{System characterization}
\label{ssec:sys-char}
The intra-cavity loss is estimated to be around $0.1\%$ giving an escape efficiency of $98.24\%$.
The system's losses consist of optical losses and losses due to tapping, both for the input of the displacement beam and tapping for the cavity locks. 

\begin{table}[ht]
\begin{tabular}{ccc}
\toprule\toprule
Component                       & OPO1 Losses & OPO 2 Losses \\ 
\midrule
OPO escape efficiency           & 1.8\,\%    & 1.8\,\%       \\
Collimating Lens                & 1.1\,\%      & 0.5\,\%       \\
polarization Beam Splitter      & -           & 1.9\,\%     \\
polarization Beam Splitter      & 1.2\,\%     & 1.2\,\%       \\
Displacement Tapping            & 3.7\,\%     & 2.5\,\%      \\
polarization Beam Splitter (HD) & 0.2\,\%    & 0.3\,\%        \\ \hline
Total                           & 7.8\,\%     & 7.9\,\% \\
\bottomrule\bottomrule
\end{tabular}
\caption{Optical losses across OPOs path.}
\label{tab: optical_losses}
\end{table}

The optical losses are mainly due to reflections from surfaces of components in the beam paths such as waveplates and lenses. We summarize the results of characterization in Table \ref{tab: optical_losses}. The total loss measured from the tapping mirror of the OPO cavity to the homodyne detector diodes is  $\SI{7.56}{\%}$ and $\SI{7.69}{\%}$ for both paths. The total efficiency of the system including all sources is summarized in Table \ref{tab:toteff}, and we estimate this to be around $79.5\,\%$. This total efficiency indicates a maximal achievable squeezing of $-6.88$ dB. Considering the effect of anti-squeezing ($+17$ dB) cross-talking into the squeezing due to the noise in phase locks (standard deviation $2.5^\circ$), the maximal locked squeezing for the setup would be $-5.2$ dB, aligning well with the observed squeezing level. 

\begin{table}[t]
    \centering
    \begin{tabular}{cc}
    \toprule\toprule
        Source & Efficiency\\
    \midrule
        Optical component &  92.1\,\%\\
        LO interference efficiency  & 92.2\,\% \\
        Cavity escape efficiency & 98.3\,\% \\
        Photodiodes efficiency & 95.0\,\% \\ 
    \midrule
       Total efficiency & 79.3\,\% \\
    \bottomrule\bottomrule
    \end{tabular}
    \caption{Efficiency budget throughout the whole set-up. The efficiency from interference is calculated according to an average fringe visibility of 98.0\% at the homodyne detectors.}
    \label{tab:toteff}
\end{table}

\section{Theory}

\subsection{Preliminaries}
The object of interest in the learning task here is the random multi-time displacement process. Such a process $\Lambda$ can be fully characterized, either by a probability distribution $p(\alpha)$ or a characteristic function $\lambda(\beta)$. 

\begin{definition}[Probability distribution of a process]\label{def:chn-prob}
    The effect of the $n$-mode process on a specific input state $\hat{\rho}$, expressed using the probability distribution of the process, is: 
    \begin{align}
        \Lambda(\hat{\rho})=\int d^{2n}\alpha\, p(\alpha)D(\alpha)\hat{\rho} D^\dagger(\alpha),
    \end{align}
    where $\alpha \in \mathbb{C}^n$ is the complex array value of the $n$-displacement vector. The displacement operator acts on all the modes as $D(\alpha) := \bigotimes_{i=1}^n D(\alpha_i)$, with $D(\alpha_i):=\exp(\alpha_i \hat{a}_i^\dagger-\alpha_i^*\hat{a}_i)$ being the displacement operator for the $i$-th mode. 
\end{definition}

\begin{definition}[Characteristic function]\label{def:chn-char}
    The characteristic function and probability distribution are the Fourier transforms of one and the other and can be written as: 
    \begin{equation}
        \begin{split}
            \lambda(\beta)&=\int d^{2n}\alpha~p(\alpha)e^{\alpha^\dagger\beta-\beta^\dagger\alpha}, \\
            p(\alpha)&=\frac{1}{\pi^{2n}}\int d^{2n}\beta~\lambda(\beta)e^{\beta^\dagger\alpha-\alpha^\dagger\beta}. \\
        \end{split}
    \end{equation}
    Here, $\beta \in \mathbb{C}^n$ is the complex value of the $n$-displacement vector in Fourier space. 
\end{definition}

\begin{definition}[Quadrature operators]\label{def:quad-optr}
    The quadrature operators are defined using the creation and annihilation operators. Throughout this work, we use the convention of $\hbar=1$. The $x$- and $p$-quadrature operators can be expressed, accordingly, as:
    \begin{align}
        \hat{X}=\frac{1}{\sqrt{2}} (\hat{a}+\hat{a}^\dagger), \quad \hat{P}=\frac{1}{\sqrt{2}\,i} (\hat{a}-\hat{a}^\dagger).
    \end{align}
\end{definition}
To reveal the effect of displacement, we can measure the shift of the quadrature values of a probe state using homodyne detection. In the Heisenberg picture, the displacement would transform quadrature operators as:
\begin{alignat}{3}
    \hat{X}&\longrightarrow D^\dagger(\alpha)&\hat{X}D(\alpha) &= \hat{X} + \sqrt{2}\,\Re&(\alpha), \nonumber\\
    \hat{P}&\longrightarrow D^\dagger(\alpha)&\hat{P}D(\alpha) &= \hat{P} + \sqrt{2}\,\Im&(\alpha).
\end{alignat}
The shift of the $x$- ($p$-)quadrature value of a probe state thus provides information of the probability distribution of the process for $\alpha$ real (imaginary).

\subsection{Learning displacement processes}

To learn the properties of an $n$-mode displacement process, we use a similar protocol as the entanglement-assisted scheme in Ref.~\cite{Oh2024}. The scheme uses $n$ TMSV states. From each of the TMSV states, one of the modes is taken out and used as a mode in the $n$-mode probe state, while the other mode is regarded as a mode in the memory state. The probe state then goes through the displacement channel while the memory state remains unchanged. Then, we apply a Bell measurement on every pair of modes which were initially in a TMSV state. The measurement reveals the difference (sum) of the $x$- ($p$-)quadrature values between the modes in the probe and memory states. The results are arranged in a $n$-complex vector, denoted as $\zeta \in \mathbb{C}^n, \zeta^{(i)} = x^{(i)}+ip^{(i)}$. Then, the following Theorem has been proved by some of the authors\,\cite{Oh2024}:

\begin{theorem}[Unbiased characteristic function estimator]\label{th:est}
    The following estimator $\Tilde{\lambda}$ is an unbiased estimator for the characteristic function of a random displacement channel:
    \begin{equation}
        \Tilde{\lambda}(\beta)=\frac{1}{N}\sum_{i=1}^N\exp(e^{-2r_\textnormal{eff}}|\beta|^2)e^{\zeta^{(i)\dagger}\beta - \beta^{\dagger}\zeta^{(i)}}
        \label{eq:estimator}
    \end{equation}
    where $N$ represents the number of acquired samples of $n$-complex displacement, and \begin{align}
    r_\textnormal{eff} = -\frac{1}{2}\log\left(e^{-2r}+\frac{1-T_a}{T_a}\right)
    \end{align}
    is the effective squeezing parameter, determined by the squeezing level $r$ of the TMSV states at the point of displacement and the transmissivity $T_a$ of the setup between displacement and Bell measurement.
\end{theorem}

Further, using the Hoeffding inequality, it is also proved in Ref.~\cite{Oh2024} that the sampling complexity of the estimator giving an $\epsilon$ additive error, $|\tilde{\lambda}(\beta)-\lambda(\beta)|\leq \epsilon$ with probability at least $1-\delta$ is upper-bounded as:
\begin{align}\label{eq:upper_bound}
    N\leq 8e^{2e^{-2r_\text{eff}}|\beta|^2}\epsilon^{-2}\log 4\delta^{-1}.
\end{align}

\subsection{Lower bound on sample complexity}
In this subsection, we present lower bounds on the number of experiments needed to learn a random displacement process.
More specifically, we first review the lower bound derived in Ref.~\cite{Oh2024} for a certain family of entanglement-free learning schemes and further generalize the learning schemes to make them applicable to a multi-time displacement process consistent with the conducted experimental setup.

The family of entanglement-free learning schemes considered in Ref.~\cite{Oh2024} is illustrated in Fig.~\ref{fig-sm:entanglement_free}.
In this family of learning schemes, to learn an $m$-mode random displacement channel, a prepared probe state, possibly entangled within the associated $m$ modes, passes through a given random displacement channel and then is destructively measured in a certain measurement basis to obtain information about the channel from the measurement outcome.
One is then allowed to choose the probe state and the measurement basis for subsequent rounds adaptively to optimize the learning scheme.
For these entanglement-free learning schemes, Ref.~\cite{Oh2024} presented a lower bound of sampling complexity for learning the characteristic function of a random displacement channel, which is captured by the following theorem:

\begin{figure}
    \centering
    \includegraphics[width=0.9\linewidth]{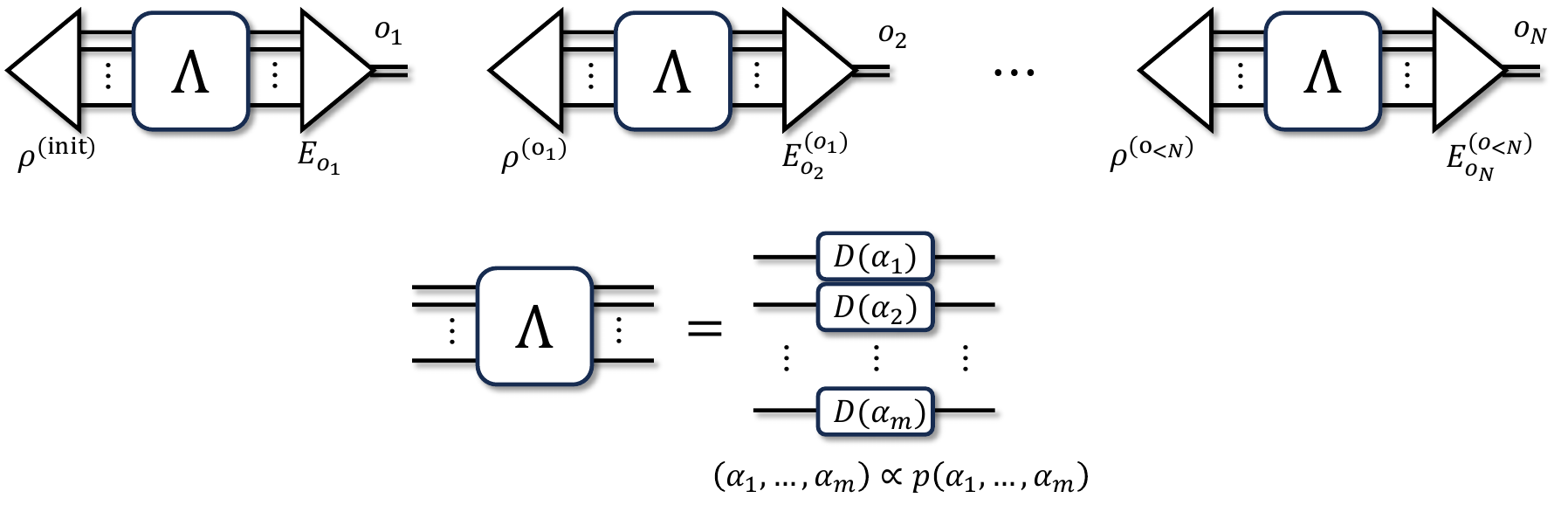}
    \caption{Schematics for entanglement-free schemes. Here, an $m$-mode random displacement channel $\Lambda$ is written as a product of displacements by $\alpha_1,\dots,\alpha_m$ for each mode, which are correlated by the underlying probability distribution $p(\alpha_1,\dots,\alpha_m)$.}
    \label{fig-sm:entanglement_free}
\end{figure}

\begin{theorem}\label{th:main}
        Let $\Lambda$ be an arbitrary $m$-mode random displacement channel ($m\ge 8$) and consider an entanglement-free scheme that uses $N$ copies of $\Lambda$. After all measurements are completed, the scheme receives the query $\beta\in \mathbb{C}^m$ and returns an estimate $\tilde\lambda(\beta)$ of $\Lambda$'s characteristic function $\lambda(\beta)$. Suppose that, with success probability at least $2/3$, $|\tilde{\lambda}(\beta)-\lambda(\beta)|\le \epsilon\le 0.24$ for all $\beta$ such that $|\beta|^2 \le \kappa m$. Then $N\ge 0.01\epsilon^{-2}(1+1.98\kappa)^{m}$.
\end{theorem}
\noindent This lower bound on $N$ in entanglement-free schemes far exceeds the upper bound Eq.~\eqref{eq:upper_bound} on $N$ in entanglement-enhanced schemes when $|\beta|^2\leq \kappa m$ and the effective squeezing parameter $r_\text{eff}$ is sufficiently large. 

\begin{figure}
    \centering
    \includegraphics[width=0.9\linewidth]{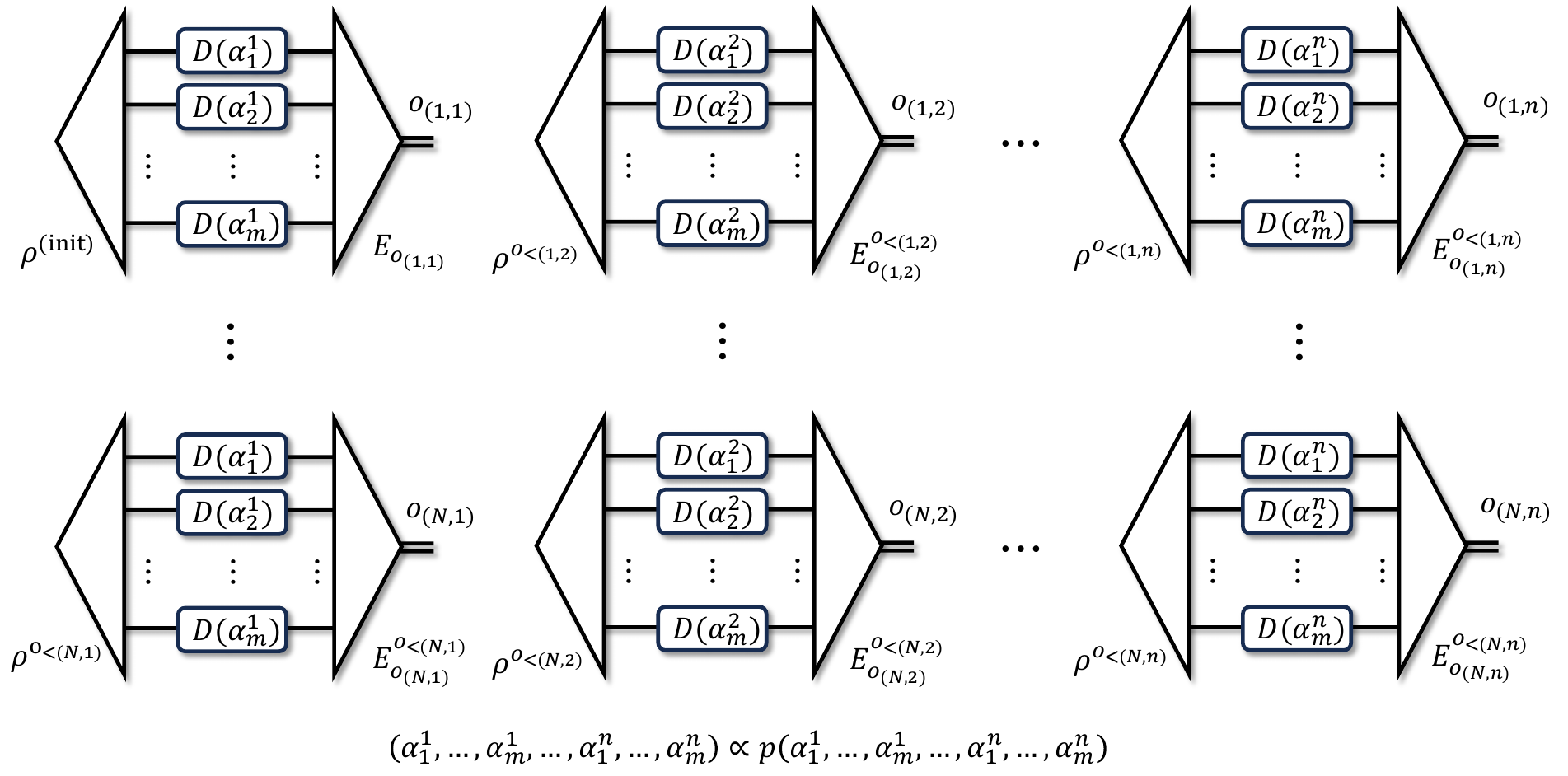}
    \caption{Schematics for entanglement-free schemes that allow adaptive strategies within a random displacement process.
    Here, a random displacement process is written as a product of displacements by $\alpha_1^1,\dots,\alpha_m^1,\dots,\alpha_1^n,\dots,\alpha_m^n$ for each mode, which are correlated by the underlying probability distribution $p(\alpha_1^1,\dots,\alpha_m^1,\dots,\alpha_1^n,\dots,\alpha_m^n)$. The $mn$ modes are grouped into $n$ bins, with each bin containing $m$ modes. One is allowed to choose the measurement basis for each bin adaptively, conditioned on the outcomes of measurements of modes in previous bins. 
    }
    \label{fig-sm:entanglement_free2}
\end{figure}

On the other hand, in the experiment, we implement a random displacement process on $n$ modes defined in the temporal domain. Because the displacement process acts on modes that are in different time bins, the probability distribution $p(\alpha)$ characterizing the process captures correlations in in time. Such temporal correlations, which were not considered in Ref.~\cite{Oh2024}, allow different learning schemes which are still entanglement free but in which the measurement basis for modes at later times can be chosen adaptively conditioned on measurement outcomes for modes at earlier times,  as illustrated in Fig.~\ref{fig-sm:entanglement_free2}.

We denote such a multi-time process as an $(m,n)$-mode random displacement process~\cite{chiribella2008quantum}, meaning that there are $n$ different time bins with $m$ modes defined in each of these time bins. Thus the case $n=1$ was considered in Ref.~\cite{Oh2024} and the entanglement-free version of our experiment realizes the case $m=1$.
Since the $mn$ single-mode displacement operators are sampled from a joint distribution $p(\alpha)$, we can define its characteristic function $\lambda(\beta)$ in the same way as for an $nm$-mode random displacement channel. 
Fortunately, a lower bound on sample complexity of entanglement-free schemes similar to \autoref{th:main} also applies to such multi-time displacement processes.
\begin{theorem}\label{th:main2}
        Let $\Lambda$ be an arbitrary $(m,n)$-mode random displacement process ($mn\ge 8$) and consider an entanglement-free scheme that uses $N$ copies of $\Lambda$. After all measurements are completed, the scheme receives the query $\beta\in \mathbb{C}^{mn}$ and returns an estimate $\tilde\lambda(\beta)$ of $\Lambda$'s characteristic function $\lambda(\beta)$. Suppose that, with success probability at least 2/3, $|\tilde{\lambda}(\beta)-\lambda(\beta)|\le \epsilon\le 0.24$ for all $\beta$ such that $|\beta|^2 \le \kappa mn$. Then $N\ge 0.01\epsilon^{-2}(1+1.98\kappa)^{mn}$.
\end{theorem}
We emphasize that we do not constrain anything such as the energy of the input state or measurement in this lower bound.
To prove \autoref{th:main2}, we utilize two families of processes---the ``three-peak'' random displacement process and the Gaussian random displacement process. We first prove that learning this restricted family of processes is hard; i.e., the lower bound on the sampling complexity scales exponentially with $mn$. 

\begin{definition}[three-peak process]\label{def:3-peak-chn}
     The family of ``three-peak'' random displacement processes, $\Lambda_\gamma$, is defined by their characteristic functions:
    \begin{align}\label{eq:lower_instance2}
        \Lambda_\gamma:~\lambda_{\gamma}(\beta)
        \equiv 
        e^{-\frac{|\beta|^2}{2\sigma^2}}+2i\epsilon_0 e^{-\frac{|\beta-\gamma|^2}{2\sigma^2}}-2i\epsilon_0 e^{-\frac{|\beta+\gamma|^2}{2\sigma^2}},\quad \gamma\in\mathbb C^{mn},
    \end{align}
    where $\epsilon_0 \equiv \epsilon/0.98 \le0.25$. We assume that $mn\ge8$ and $\epsilon\leq 0.24$.
    The set of all three-peak processes with parameters $(\epsilon,\sigma)$ is denoted as $\bm\Lambda^{\epsilon,\sigma}_\text{3-peak}$. Note that narrow peaks (small $\sigma$) in the characteristic function $\lambda(\beta)$ correspond via the Fourier transform to a probability distribution $p(\alpha)$ supported on large displacements.
\end{definition}

\begin{definition}[Gaussian process]\label{def:gaussian-chn}
     The Gaussian random displacement process is defined by its characteristic function:
    \begin{align}\label{eq:lower_instance2}
        \Lambda_0:~\lambda(\beta)
        \equiv 
        e^{-\frac{|\beta|^2}{2\sigma^2}},
    \end{align}
    The subscript in the notation is because the Gaussian process can be thought of as an extreme case of the three-peak process where $|\gamma|\to 0$.
\end{definition}

We now prove that even if one knows that a given process to be learned is from the restricted family, $\bm\Lambda^{\epsilon,\sigma}_\text{3-peak}$, an exponential lower bound still applies.
This leads to Theorem~\ref{th:main2} as a corollary.

\begin{theorem}\label{th:main3}
        Given positive numbers $m, n, \sigma, \kappa, \epsilon$ such that
    \begin{equation}\label{eq:thS2_cond}
        2\sigma^2 \le \max\left\{1-1.98\kappa,~0.99\kappa\left(\sqrt{1+{(0.99\kappa)^{-2}}}-1\right)\right\}, \quad mn\ge8 , \quad \epsilon\le0.24.
    \end{equation}
    If there exists an entanglement-free scheme such that, after learning from $N$ copies of an $(m,n)$-mode random displacement process $\Lambda\in\bm\Lambda^{\epsilon,\sigma}_\textnormal{3-peak}$, and then receiving a query $\beta\in\mathbb{C}^{mn}$, can return an estimate $\tilde\lambda(\beta)$ of $\lambda(\beta)$ such that $|\tilde\lambda(\beta)-\lambda(\beta)|\le\epsilon$ with probability at least $2/3$ for all $\beta$ such that $|\beta|^2\le \kappa mn$, then 
    \begin{equation}
        N\geq 0.01\epsilon^{-2} \left(1+\frac{1.98\kappa}{1+2\sigma^2}\right)^{mn}.
    \end{equation}
\end{theorem}
\noindent By setting $\sigma\to0$, we obtain Theorem~\ref{th:main2}.

Now, we prove the theorem, which follows a similar procedure to that of Theorem~\ref{th:main}.
Let us define some notations first. Let $\rho^{\bm{o}_{<(k,t)}}$ and $\{E^{\bm{o}_{<(k,t)}}_{o_{(k,t)}}\}_{o_{(k,t)}}$ denote the input state at step $(k,t)$  depending on the previous measurement outcomes $\bm{o}_{\leq (k,t)}$ and the measurement POVM elements with measurement outcome $o_{(k,t)}$, as described in Fig.\,\ref{fig-sm:entanglement_free2}. Additionally, we define $\bm{o}_{(k,:)}$ to denote the list of outcomes $[o_{(k,1)},\dots,o_{(k,n)}]$ and $\bm{o}$ to be the list of all outcomes.

We define a hypothesis-testing game to prove the lower bound. 
A referee samples a tuple $(\gamma,s)$ such that $\gamma\in \mathbb{C}^{mn}$ is from a normal distribution with variance $\sigma_\gamma^2\equiv 0.99\kappa/2$ and $s\in\{\pm 1\}$ is from a uniform distribution. Then the referee chooses one of the following random displacement processes with equal probability,
\begin{itemize}
    \item $\Lambda_0:\lambda_0(\beta)=e^{-\frac{|\beta|^2}{2\sigma^2}}$,
    \item $\Lambda_{\gamma,s}:\lambda_{\gamma,s}(\beta)=e^{-\frac{|\beta|^2}{2\sigma^2}}+2is\epsilon_0\left(e^{-\frac{|\beta-s\gamma|^2}{2\sigma^2}}-e^{-\frac{|\beta+s\gamma|^2}{2\sigma^2}}\right)\equiv \lambda_0(\beta)+\lambda_{s\gamma}^{\text{add}}(\beta)$.
\end{itemize}
The player is then asked to learn from $N$ copies of the chosen $\Lambda$. After the learning is completed, the referee reveals the value of $\gamma$ (but not $s$) to the player, who is then challenged to decide whether the referee has chosen $\Lambda_0$ or $\Lambda_{\gamma,s}$.

Suppose that there exists a learning scheme with $N$ copies that satisfies the assumption of \autoref{th:main2}; i.e., such that the characteristic function $\lambda(\beta)$ can be estimated within additive error $\epsilon$ with probability at least $2/3$ for $\beta$ such that $|\beta|^2\leq \kappa mn$. Then the player can use this scheme in the hypothesis-testing game. If the revealed value of $\gamma$ satisfies $2\sigma^2<|\gamma|^2\leq \kappa mn$, the player evaluates $\tilde \lambda(\gamma)$ and chooses the process such that $\lambda(\gamma)$ most closely matches this value; otherwise, the player randomly guesses the process chosen by the referee.

The effectiveness of this strategy for playing the hypothesis-testing game was analyzed in Ref.~\cite{Oh2024}, where further details can be found. One shows that $\Pr(2\sigma^2<|\gamma|^2\leq \kappa mn)\geq 0.49987$ when $mn\geq 8$ and $2\sigma^2\leq 0.99\kappa$. Furthermore, when $|\gamma|^2$ lies in this range, $|\lambda_0(\gamma)-\lambda_{\gamma,s}(\lambda)|\ge 2\epsilon$, and thus the player's success probability is lower-bounded by $0.49987 * 2/3 + (1-0.49987) * 1/2 \ge 0.5833$. 
Therefore, using the relation between the success probability of the hypothesis testing and the total variation distance,
\begin{align}
    \Pr[\text{Success}]=\mathbb{E}_{\gamma}\Pr[\text{Success}|\gamma]
    \leq \frac{1}{2}\left(1+\mathbb{E}_\gamma \text{TVD}(\Pr_{\gamma=0}[\bm{o}],\mathbb{E}_s\Pr_{\gamma,s}[\bm{o}])\right),
    \label{eq:prob_tvd}
\end{align}
we have that
\begin{align}\label{eq:success-TVD}
    \mathbb{E}_\gamma \text{TVD}(\Pr_{\gamma=0}[\bm{o}],\mathbb{E}_s\Pr_{\gamma,s}[\bm{o}])\geq 0.1666.
\end{align}
We now obtain an upper bound on the left-hand side and show that 
Eq.~\eqref{eq:success-TVD} cannot be satisfied unless the number of samples $N$ is exponentially large in $mn$.

First of all, the output probability is written as 
\begin{align}
    \Pr[\bm{o}]
    =\Pr[\bm{o}_{(1,:)}]\Pr[\bm{o}_{(2,:)}|\bm{o}_{<(2,1)}]\cdots \Pr[\bm{o}_{(N,:)}|\bm{o}_{<(N,1)}],
\end{align}
where the conditional probability is
\begin{align}
    \Pr[\bm{o}_{(k,:)}|\bm{o}_{<(k,1)}]
    &=\int d^{2mn}\alpha~p(\alpha)\Pr[\bm{o}_{(k,:)}|\bm{o}_{<(k,1)},\alpha] \\
    &=\int d^{2mn}\alpha~p(\alpha)\prod_{t=1}^n \Pr[\bm{o}_{(k,t)}|\bm{o}_{<(k,t)},\alpha] \\ 
    &=\int d^{2mn}\alpha~p(\alpha)\prod_{t=1}^n \Tr[E_{\bm{o}_{k,t}}^{\bm{o}_{<(k,t)}}D(\alpha_t)\rho^{\bm{o}_{<(k,t)}}D^\dagger (\alpha_t)] \\ 
    &=\int d^{2mn}\alpha~p(\alpha)\Tr\left[\bigotimes_{t=1}^n E_{\bm{o}_{k,t}}^{\bm{o}_{<(k,t)}} \bigotimes_{t=1}^n D(\alpha_t)\bigotimes_{t=1}^n \rho^{\bm{o}_{<(k,t)}}\bigotimes_{t=1}^n D^\dagger (\alpha_t)\right] \\ 
    &=\Tr\left[\bigotimes_{t=1}^n E_{\bm{o}_{k,t}}^{\bm{o}_{<(k,t)}} \Lambda\left(\bigotimes_{t=1}^n \rho^{\bm{o}_{<(k,t)}}\right)\right],
\end{align}
where $\alpha\in \mathbb{C}^{mn}$, $\alpha^t\in \mathbb{C}^{m}$. Here $\Lambda$ is an $mn$-mode random displacement process with the same characteristic function as the $(m,n)$-mode random displacement process to be learned. Note that,
\begin{align}
    \Lambda(\rho)
    &=\int d^{2mn}\alpha~p(\alpha) \bigotimes_{t=1}^n D(\alpha^t)\rho \bigotimes_{t=1}^n D^\dagger(\alpha_t) \\
    &=\frac{1}{\pi^{mn}}\int d^{2mn}\alpha \prod_{t=1}^n(d^{2m}\beta^t)~p(\alpha) \Tr\left[\rho \bigotimes_{t=1}^{n} D(\beta^t)\right]\bigotimes_{t=1}^n D(\alpha^t)\bigotimes_{t=1}^{n} D^\dagger(\beta^t) \bigotimes_{t=1}^n D^\dagger(\alpha^t) \\ 
    &=\frac{1}{\pi^{mn}}\int d^{2mn}\alpha d^{2mn}\beta~p(\alpha) \Tr\left[\rho \bigotimes_{t=1}^{n} D(\beta^t)\right]\bigotimes_{t=1}^{n} D^\dagger(\beta^t) \prod_{t=1}^n e^{\beta^{t\dagger}\alpha^{t}-\alpha^{t\dagger}\beta^{t}} \\ 
    &=\frac{1}{\pi^{mn}}\int d^{2mn}\beta \lambda(\beta) \Tr\left[\rho \bigotimes_{t=1}^{n} D(\beta^t)\right]\left[\bigotimes_{t=1}^{n} D^\dagger(\beta^t)\right].
\end{align}
Using this expression, we simplify the difference of the associated probability distributions to compute the total variation distance as
\begin{align}
    &\Pr_{\gamma=0}[\bm{o}]-\mathbb{E}_s\Pr_{\gamma,s}[\bm{o}] \\
    &=\Pr_{\gamma=0}[\bm{o}]\left(1-\mathbb{E}_s\frac{\Pr_{\gamma,s}[\bm{o}]}{\Pr_{\gamma=0}[\bm{o}]}\right) \\ 
    &=\Pr_{\gamma=0}[\bm{o}]\left(1-\mathbb{E}_s \prod_{k=1}^N \frac{\Pr_{\gamma,s}[\bm{o}_{(k,:)}|\bm{o}_{<(k,1)}]}{\Pr_{\gamma=0}[\bm{o}_{(k,:)}|\bm{o}_{<(k,1)}]}\right) \\ 
    &=\Pr_{\gamma=0}[\bm{o}]\left(1-\mathbb{E}_s \prod_{k=1}^N \frac{\Tr[\bigotimes_{t=1}^n E_{\bm{o}_{k,t}}^{\bm{o}_{<(k,t)}} \Lambda_{s\gamma}\left(\bigotimes_{t=1}^n \rho^{\bm{o}_{<(k,t)}}\right)]}{\Tr[\bigotimes_{t=1}^n E_{\bm{o}_{k,t}}^{\bm{o}_{<(k,t)}} \Lambda_{0}\left(\bigotimes_{t=1}^n \rho^{\bm{o}_{<(k,t)}}\right)]}\right) \\ 
    &=\Pr_{\gamma=0}[\bm{o}]\left(1-\mathbb{E}_s \prod_{k=1}^N \left(1+s\frac{\Tr[\bigotimes_{t=1}^n E_{\bm{o}_{k,t}}^{\bm{o}_{<(k,t)}} \Lambda^{\text{add}}_{\gamma}\left(\bigotimes_{t=1}^n \rho^{\bm{o}_{<(k,t)}}\right)]}{\Tr[\bigotimes_{t=1}^n E_{\bm{o}_{k,t}}^{\bm{o}_{<(k,t)}} \Lambda_{0}\left(\bigotimes_{t=1}^n \rho^{\bm{o}_{<(k,t)}}\right)]}\right)\right) \\ 
    &=\Pr_{\gamma=0}[\bm{o}]\left(1-\mathbb{E}_s\prod_{k=1}^N\left(\frac{\frac{1}{\pi^{mn}}\int d^{2mn}\beta\lambda_{\gamma,\text{add}}(\beta)\prod_{t=1}^n\left(\Tr[E_{\bm{o}_{k,t}}^{\bm{o}_{<(k,t)}}D^\dagger(\beta^t)]\Tr[\rho^{\bm{o}_{<(k,t)}}D(\beta^t)]\right)}{\frac{1}{\pi^{mn}}\int d^{2mn}\beta\lambda_0(\beta)\prod_{t=1}^n\left(\Tr[E_{\bm{o}_{k,t}}^{\bm{o}_{<(k,t)}}D^\dagger(\beta^t)]\Tr[\rho^{\bm{o}_{<(k,t)}}D(\beta^t)]\right)}\right)\right) \\ 
    &=\Pr_{\gamma=0}[\bm{o}]\left(1-\mathbb{E}_s \prod_{k=1}^N \left(1-4s\epsilon_0 \text{Im}~G_{s\gamma}^{\bm{o}_{\leq (k,n)}}\right)\right),
\end{align}
where we defined
\begin{align}
    G_\gamma^{\bm{o}_{\leq (k,n)}}
    \equiv \frac{\int d^{2mn}\beta e^{-\frac{|\beta-\gamma|^2}{2\sigma^2}}G^{\bm{o}_{\leq (k,n)}}(\beta)}{\int d^{2mn}\beta e^{-\frac{|\beta|^2}{2\sigma^2}}G^{\bm{o}_{\leq (k,n)}}(\beta)},
\end{align}
and
\begin{align}
    G^{\bm{o}_{\leq (k,n)}}(\beta)
    &\equiv \prod_{t=1}^n \Tr[E_{\bm{o}_{k,t}}^{\bm{o}_{<(k,t)}}D^\dagger(\beta^t)]\Tr[\rho^{\bm{o}_{<(k,t)}}D(\beta^t)]\equiv \Tr[E^{k} D^\dagger(\beta)]\Tr[\rho^{k}D(\beta)],
\end{align}
where we defined $E^{k}\equiv \bigotimes_{t=1}^n E_{\bm{o}_{k,t}}^{\bm{o}_{<(k,t)}}, \rho^{k}\equiv \bigotimes_{t=1}^n \rho^{\bm{o}_{<(k,t)}}$.
Here, because $\left(1-4s\epsilon_0 \text{Im}~G_{s\gamma}^{\bm{o}_{\leq (k,n)}}\right)$ is a probability ratio, it is nonnegative.
Thus,
\begin{align}
    \mathbb{E}_{s=\pm 1} \prod_{k=1}^N \left(1-4s\epsilon_0 \text{Im}~ G_{\gamma}^{\bm{o}_{\leq (k,n)}}\right)
    &\geq \sqrt{\prod_{k=1}^N \left(1-4\epsilon_0 \text{Im}~ G_{\gamma}^{\bm{o}_{\leq (k,n)}}\right)\left(1+4\epsilon_0 \text{Im}~G_{-\gamma}^{\bm{o}_{\leq (k,n)}}\right)} \\ 
    &= \sqrt{\prod_{k=1}^N \left(1-16\epsilon_0^2 \left(\text{Im}~ G_{\gamma}^{\bm{o}_{\leq (k,n)}}\right)^2\right)} \\ 
    &\geq \prod_{k=1}^N \left(1-16\epsilon_0^2 \left(\text{Im}~ G_{\gamma}^{\bm{o}_{\leq (k,n)}}\right)^2\right) \\
    &\geq 1-16\epsilon_0^2 \sum_{k=1}^N |G_{\gamma}^{\bm{o}_{\leq (k,n)}}|^2,
\end{align}
where the first line uses the AM-GM inequality and the fact that each term is nonnegative and the second line uses the fact that $\text{Im}~G_\gamma^{\bm{o}_{\leq (k,n)}}=-\text{Im}~G_{-\gamma}^{\bm{o}_{\leq (k,n)}}$, the third line uses $\sqrt{1-x}\geq 1-x$ for $x\in [0,1]$, and the last line uses the inequality $\prod_i(1-x_i)\geq 1-\sum_i x_i$ for all $x_i\in [0,1]$.

Therefore, we can get rid of the maximum in the expression of the average TVD. 
Thus, the average TVD is written as
\begin{align}
    \mathbb{E}_{\gamma}\text{TVD}(\Pr_{\gamma=0}[\bm{o}],\mathbb{E}_s \Pr_{\gamma,s}[\bm{o}])
    =\mathbb{E}_{\gamma}\sum_{\bm{o}}\max(0,\Pr_{\gamma=0}[\bm{o}]-\mathbb{E}_s \Pr_{\gamma,s}[\bm{o}])
    \leq \sum_{\bm{o}}\Pr_{\gamma=0}[\bm{o}] \sum_{k=1}^N 16\epsilon_0^2 \mathbb{E}_\gamma |G_{\gamma}^{\bm{o}_{\leq (k,n)}}|^2.
\end{align}
We claim that when $\sigma^2\leq \max\left\{\frac{1}{2}-2\sigma_\gamma^2,\sigma_\gamma^2\left(\sqrt{1+\frac{1}{4\sigma_\gamma^4}}-1\right)\right\}$,
\begin{align}
    \mathbb{E}_\gamma |G_{\gamma}^{\bm{o}_{\leq (k,n)}}|^2\leq \left(\frac{1+2\sigma^2}{1+2\sigma^2+4\sigma_\gamma^2}\right)^{mn}.
\end{align}
Then, the average TVD is upper-bounded by
\begin{align}
    \mathbb{E}_{\gamma}\text{TVD}(\Pr_{\gamma=0}[\bm{o}],\mathbb{E}_s \Pr_{\gamma,s}[\bm{o}])
    =\mathbb{E}_{\gamma}\sum_{\bm{o}}\max(0,\Pr_{\gamma=0}[\bm{o}]-\mathbb{E}_s \Pr_{\gamma,s}[\bm{o}])
    \leq 16N\epsilon_0^2 \left(\frac{1+2\sigma^2}{1+2\sigma^2+4\sigma_\gamma^2}\right)^{mn},
    \label{eq:ub_tvd}
\end{align}
and thus by substituting $\epsilon=0.98\epsilon_0$ and $2\sigma_\gamma^2=0.99\kappa$, the number of copies we are required to accomplish the learning task is lower-bounded as
\begin{align}
    N\geq 0.01\epsilon^{-2}\left(1+\frac{1.98\kappa}{1+2\sigma^2}\right)^{mn}.
\end{align}
which proves the theorem.

\begin{figure}
    \centering
    \includegraphics[width=0.35\linewidth]{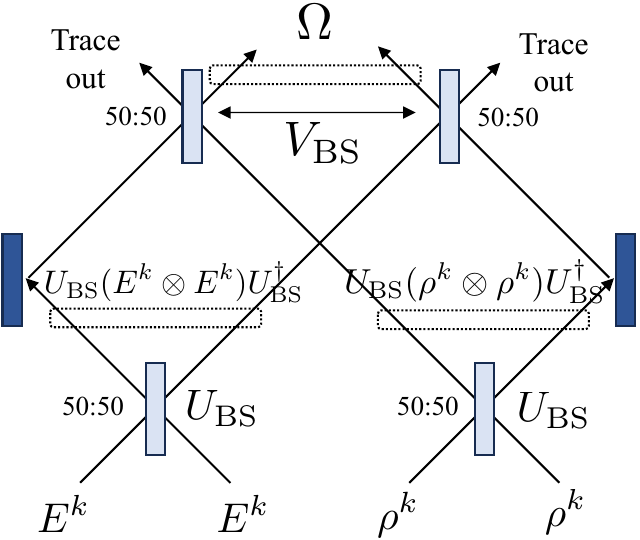}
    \caption{Schematics of Eqs.~\eqref{sm_eq:proof_first}-\eqref{sm_eq:proof_last}. Here, each line represents $n$-mode system and $V_\text{BS}$ corresponds to 50:50 beam splitters including both sides.}
    \label{fig-sm:proof}
\end{figure}

Thus, the remaining task is to prove the claimed upper bound of $\mathbb{E}_\gamma |G_{\gamma}^{\bm{o}_{\leq (k,n)}}|^2$.
First, after taking the expectation value over $\gamma$, we obtain
\begin{align}
    \mathbb{E}_\gamma |G_\gamma^{\bm{o}_{\leq (k,n)}}|^2
    =\frac{1}{\left(\frac{2\sigma_\gamma^2}{\sigma^2}+1\right)^{mn}}\frac{\int d^{2mn}\beta d^{2mn}\tilde{\beta} e^{-\frac{|\beta|^2+|\tilde{\beta}|^2}{2(\sigma^2+2\sigma_\gamma^2)}}e^{-\frac{|\beta-\tilde{\beta}|^2}{2\sigma^2(\sigma^2+2\sigma_\gamma^2)/\sigma_\gamma^2}}G^{\bm{o}_{\leq (k,n)}*}(\beta)G^{\bm{o}_{\leq (k,n)}}(\tilde{\beta})}{\int d^{2mn}\beta d^{2mn}\tilde{\beta} e^{-\frac{|\beta|^2+|\tilde{\beta}|^2}{2\sigma^2}}G^{\bm{o}_{\leq (k,n)}*}(\beta)G^{\bm{o}_{\leq (k,n)}}(\tilde{\beta})}.
\end{align}
Here, the common factor $G^{\bm{o}_{\leq (k,n)}*}(\beta)G^{\bm{o}_{\leq (k,n)}}(\tilde{\beta})$ can be simplified as
\begin{align}
    G^{\bm{o}_{\leq (k,n)}*}(\beta)G^{\bm{o}_{\leq (k,n)}}(\tilde{\beta})
    &=\Tr[(E^{k}\otimes E^{k})(D(\beta)\otimes D^\dagger(\tilde{\beta}))]\Tr[(\rho^{k}\otimes \rho^{k})(D^\dagger(\beta)\otimes D(\tilde{\beta}))] \\ 
    &=\chi_{E^{k}\otimes E^{k}}(\beta,-\tilde{\beta})\chi_{\rho^{k}\otimes \rho^{k}}(-\beta,\tilde{\beta}) \label{sm_eq:proof_first}\\ 
    &=\chi_{E^{k}\otimes E^{k}}\left(\frac{\beta_++\beta_-}{2},\frac{\beta_--\beta_+}{2}\right)\chi_{\rho^{k}\otimes \rho^{k}}\left(-\frac{\beta_++\beta_-}{2},\frac{\beta_+-\beta_-}{2}\right) \\ 
    &=\chi_{U_\text{BS}(E^{k}\otimes E^{k})U_\text{BS}^\dagger}\left(\frac{\beta_+}{\sqrt{2}},\frac{\beta_-}{\sqrt{2}}\right)\chi_{U_\text{BS}(\rho^{k}\otimes \rho^{k})U_\text{BS}^\dagger}\left(-\frac{\beta_+}{\sqrt{2}},-\frac{\beta_-}{\sqrt{2}}\right) \label{sm_eq:proof_fourth} \\ 
    &=\chi_{\Omega}(\beta_+,\beta_-), \label{sm_eq:proof_last}
\end{align}
where, for the second line, we defined the characteristic function of an operator $A$ as $\chi_A(\beta)\equiv \Tr[AD(\beta)]$. 
We illustrate how the associated operators transform in the series of equations in Fig.~\ref{fig-sm:proof}.
For the third line, we defined $\beta_\pm\equiv \beta \pm \tilde{\beta}$; for the fourth line, we transformed the associated operators by $50:50$ beam splitters $U_\text{BS}$, which in turn transforms the variables accordingly. 
For the last line, we defined an operator $\Omega\equiv \Tr_2[V_\text{BS}(U_\text{BS}(E^k\otimes E^k)U_\text{BS}^\dagger\otimes U_\text{BS}(\rho^k\otimes \rho^k)U_\text{BS}^\dagger)V_\text{BS}^\dagger]$, where the label 2 for partial trace accounts for the second half of both of $U_\text{BS}(E^k\otimes E^k)U^\dagger_\text{BS}$ and $U_\text{BS}(\rho^k\otimes \rho^k)U^\dagger_\text{BS}$ after additional 50:50 beam splitters $V_\text{BS}$. 
To understand why the operator $\Omega$ appears, note that when we have two operators in a product form, followed by 50:50 beam splitters $V_\text{BS}$, their characteristic function is written as
\begin{align}
    \chi_{V_\text{BS}(A\otimes B)V_\text{BS}^\dagger}(\alpha,\beta)=\Tr[V_\text{BS}(A\otimes B)V_\text{BS}^\dagger (D(\alpha)\otimes D(\beta))]
\end{align}
and if we trace out the second part, which is equivalent to setting $\beta=0$, the characteristic function of the reduced state is given by
\begin{align}
    \chi_{\Tr_2[V_\text{BS}(A\otimes B)V_\text{BS}^\dagger]}(\alpha)
    &=\chi_{V_\text{BS}(A\otimes B)V_\text{BS}^\dagger}(\alpha,0) \\
    &=\Tr[V_\text{BS}(A\otimes B)V_\text{BS}^\dagger (D(\alpha)\otimes I)] \\
    &=\Tr[(A\otimes B)V_\text{BS}^\dagger (D(\alpha)\otimes I)V_\text{BS}] \\
    &=\Tr[(A\otimes B)(D(\alpha/\sqrt{2})\otimes D(-\alpha/\sqrt{2}))] \\ 
    &=\chi_A(\alpha/\sqrt{2})\chi_B(-\alpha/\sqrt{2}).
\end{align}
Thus, by associating $A$ and $B$ with $U_\text{BS}(E^k\otimes E^k)U_\text{BS}^\dagger$ and $U_\text{BS}(\rho^k\otimes \rho^k)U_\text{BS}^\dagger$ in Eq.~\eqref{sm_eq:proof_fourth}, respectively, we obtain the last line.

Therefore, after changing the variables $\beta,\tilde{\beta}$ to $\beta_\pm$, we have
\begin{align}
    \mathbb{E}_\gamma |G_\gamma^{\bm{o}_{\leq (k,n)}}|^2 
    &=\frac{1}{\left(\frac{2\sigma_\gamma^2}{\sigma^2}+1\right)^{mn}}\frac{\int d^{2mn}\beta_+ d^{2mn}\beta_- e^{-\frac{|\beta_+|^2}{4(\sigma^2+2\sigma_\gamma^2)}}e^{-\frac{|\beta_-|^2}{4\sigma^2}}\chi_{\Omega}(\beta_+,\beta_-)}{\int d^{2mn}\beta_+ d^{2mn}\beta_- e^{-\frac{|\beta_+|^2}{4\sigma^2}}e^{-\frac{|\beta_-|^2}{4\sigma^2}}\chi_{\Omega}(\beta_+,\beta_-)} \\ 
    &=\left(\frac{1+2\sigma^2}{1+2\sigma^2+4\sigma_\gamma^2}\right)^{mn}\frac{\Tr\left[\left[\bigotimes_{i=1}^{mn} \left(\frac{1-2(\sigma^2+2\sigma_\gamma^2)}{1+2(\sigma^2+2\sigma_\gamma^2)}\right)^{\hat{n}_i} \bigotimes_{i=1}^{mn} \left(\frac{1-2\sigma^2}{1+2\sigma^2}\right)^{\hat{n}_i}\right]\Omega\right]}{\Tr\left[\left[\bigotimes_{i=1}^{mn} \left(\frac{1-2\sigma^2}{1+2\sigma^2}\right)^{\hat{n}_i} \bigotimes_{i=1}^{mn} \left(\frac{1-2\sigma^2}{1+2\sigma^2}\right)^{\hat{n}_i}\right]\Omega\right]} \\ 
    &\leq \left(\frac{1+2\sigma^2}{1+2\sigma^2+4\sigma_\gamma^2}\right)^{mn},
\end{align}
where, for the second line, we used the relation $\Tr[AB]=\pi^{-2n}\int d^{2n}\beta \chi_A(\beta)\chi_B^*(\beta)$ for $n$-mode operators $A$ and $B$, and the following relation~\cite{ferraro2005gaussian}:
\begin{align}
    \chi_{\frac{2\sigma^2}{1+\sigma^2}\left(\frac{1-\sigma^2}{1+\sigma^2}\right)^{\hat n}}(\alpha) \equiv \frac{2\sigma^2}{1+\sigma^2}\Tr\left[D(\alpha)\left(\frac{1-\sigma^2}{1+\sigma^2}\right)^{\hat{n}}\right]=e^{-\frac{|\alpha|^2}{2\sigma^2}}.
\end{align}
Here, $\hat{n}_i$ represents the number operator of the $i$th mode.
Finally, for the last inequality, we consider two different parameter regimes.
First, if $2\sigma^2+4\sigma_\gamma^2\leq 1$, then because the operator in the parenthesis is positive-semidefinite and, by monotonicity, the operator of the numerator is smaller than or equal to that of the denominator, which gives the desired inequality.
Second, if $2\sigma^2+4\sigma_\gamma^2>1$ but $2\sigma^2\leq 2\sigma_\gamma^2\left(\sqrt{1+\frac{1}{4\sigma_\gamma^2}}-1\right)\leq 1$ (the last inequality holds when $\sigma_\gamma>0$, ), the above can be bounded as
\begin{align}
    \frac{\Tr\left[\left[\bigotimes_{i=1}^{mn} \left(\frac{1-2(\sigma^2+2\sigma_\gamma^2)}{1+2(\sigma^2+2\sigma_\gamma^2)}\right)^{\hat{n}_i} \bigotimes_{i=1}^{mn} \left(\frac{1-2\sigma^2}{1+2\sigma^2}\right)^{\hat{n}_i}\right]\Omega\right]}{\Tr\left[\left[\bigotimes_{i=1}^{mn} \left(\frac{1-2\sigma^2}{1+2\sigma^2}\right)^{\hat{n}_i} \bigotimes_{i=1}^{mn} \left(\frac{1-2\sigma^2}{1+2\sigma^2}\right)^{\hat{n}_i}\right]\Omega\right]}
    &\leq 
    \frac{\Tr\left[\left[\bigotimes_{i=1}^{mn} \left|\frac{1-2(\sigma^2+2\sigma_\gamma^2)}{1+2(\sigma^2+2\sigma_\gamma^2)}\right|^{\hat{n}_i} \bigotimes_{i=1}^{mn} \left(\frac{1-2\sigma^2}{1+2\sigma^2}\right)^{\hat{n}_i}\right]\Omega\right]}{\Tr\left[\left[\bigotimes_{i=1}^{mn} \left(\frac{1-2\sigma^2}{1+2\sigma^2}\right)^{\hat{n}_i} \bigotimes_{i=1}^{mn} \left(\frac{1-2\sigma^2}{1+2\sigma^2}\right)^{\hat{n}_i}\right]\Omega\right]} \\ 
    &=\frac{\Tr\left[\left[\bigotimes_{i=1}^{mn} \left(\frac{1-2\Sigma^2}{1+2\Sigma^2}\right)^{\hat{n}_i} \bigotimes_{i=1}^{mn} \left(\frac{1-2\sigma^2}{1+2\sigma^2}\right)^{\hat{n}_i}\right]\Omega\right]}{\Tr\left[\left[\bigotimes_{i=1}^{mn} \left(\frac{1-2\sigma^2}{1+2\sigma^2}\right)^{\hat{n}_i} \bigotimes_{i=1}^{mn} \left(\frac{1-2\sigma^2}{1+2\sigma^2}\right)^{\hat{n}_i}\right]\Omega\right]} \\
    &\leq 1,
\end{align}
where, in the second line, we define $-\frac{1-2\sigma^2-4\sigma_\gamma^2}{1+2\sigma^2+4\sigma_\gamma^2}\equiv\frac{1-2\Sigma^2}{1+2\Sigma^2}$, i.e., $\Sigma^2=\frac{1}{4(\sigma^2+2\sigma_\gamma^2)}$.
In the third line, we used $\Sigma^2\ge\sigma^2$, which can be easily verified under our assumptions for $\sigma$.
Therefore, as long as $\sigma^2 \le \max\left\{\frac12-2\sigma_\gamma^2,~\sigma_\gamma^2\left(\sqrt{1+\frac1{4\sigma_\gamma^4}}-1\right)\right\}$, the above inequality holds.

\section{Process reconstruction}
\label{sec:chn-recon}

\subsection{Problem setting}

In this section, we demonstrate the squeezing-enhanced multi-time process reconstruction based on the estimator in Eq.~\eqref{eq:estimator}. Specifically, we aim to reconstruct with good fidelity the characteristic function of an $n$-mode three-peak displacement using $n\times N$ Bell measurement on the memory and the displaced probe modes. To highlight and assess the influence of the squeezing on the fidelity of the reconstruction, we implement this learning approach for the case of a vacuum heterodyne two-mode state compared to a two-mode squeezed vacuum state, and analyze the sample complexity scaling across different numbers of modes. 

The $n$-mode displacement process $\bm\Lambda_{\text{3-peak}}^{\epsilon,\sigma}$, as defined in~\autoref{def:3-peak-chn}, is characterized by a three-peak distribution of its characteristic function. We recall the form of its characteristic function and provide its probability distribution:
\begin{equation}
    \lambda_\gamma(\beta) = e^{-\frac{|\beta|^2}{2\sigma^2}}+2i\epsilon_0e^{-\frac{|\beta-\gamma|^2}{2\sigma^2}}-2i\epsilon_0e^{-\frac{|\beta+\gamma|^2}{2\sigma^2}}
\end{equation}
\begin{equation}
    p_\gamma(\alpha)\propto e^{-2\sigma^2|\alpha|^2}\Big(1+4\epsilon_0\text{sin}\big(2\left(\gamma_i\cdot\alpha_r-\gamma_r\cdot\alpha_i\right)\big)\Big).
\end{equation}
The subscripts $r$ and $i$ denote the real and imaginary components, respectively. Throughout this part of the experiment, we fixed the 
side peaks height to be 0.5 and its location
along the main diagonal: $\forall n \in \mathbb{N^*}, \gamma = (0.3+0.3j)\cdot\openone_{n\times1}$. This results in a choice of parameters of $\sigma = 0.3$, $|\gamma| = 0.3\cdot\sqrt{2n}$, and $\epsilon_0 = 0.25$.

\subsection{Visualizing quantum improvement}

\begin{figure}[t]
    \centering
    \includegraphics[width=\linewidth]{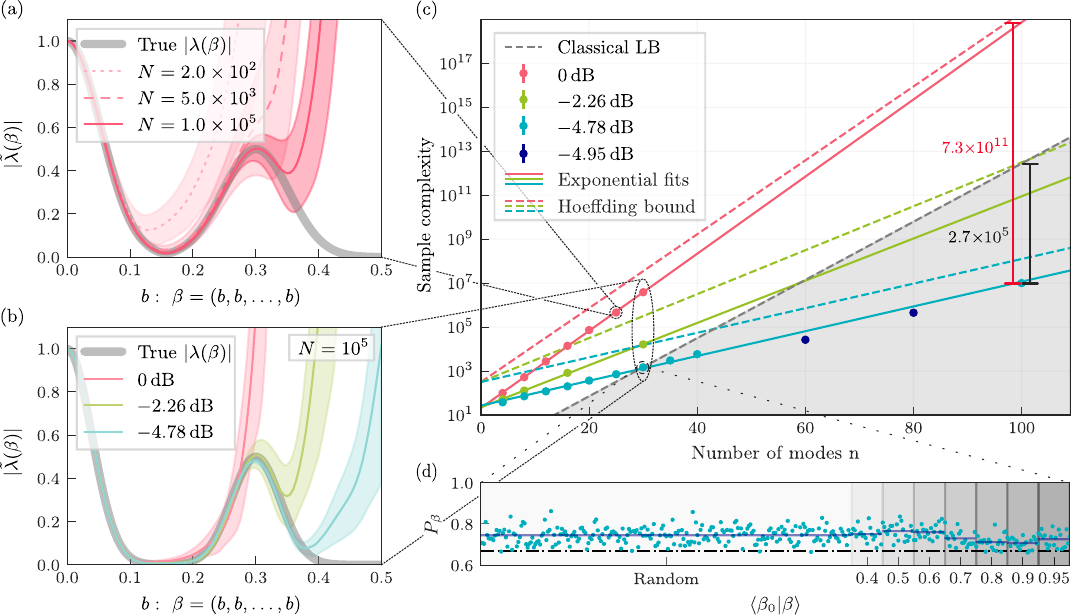}
    \caption{(a) Experimental result of the reconstructed characteristic function $\tilde{\lambda}(\beta)$ of a given process using heterodyne measurement (without squeezing) with different numbers of samples, compared with the true characteristic function $\lambda(\beta)$. 
    Here, the number of modes is $n=16$, and the plot shows a slice of the high-dimensional characteristic function space along the $\beta_0\propto(1, 1, \ldots, 1)$ direction, where the displacement locates.
    The lines show the average outcome of 100 runs of the reconstruction task, and the shades denote the $1\sigma$ standard deviation.
    (b) Same as above, but using Bell measurement (with different amounts of squeezing). Here the number of modes is $n=30$ and we always use $10^5$ samples for the same task.
    (c) Required number of samples to $\epsilon$-close reconstruct $\lambda(\beta)$ with a success probability of $1-\delta=\frac{2}{3}$ versus the number of modes along the $\beta_0$ direction. The points are calculated using experimental data and the $1\sigma$ standard deviation error bars are smaller than the data points.
    The dashed lines are Hoeffding bound calculated from Eq.\,\eqref{eq:upper_bound}, denoting the theoretical loose lower bound from which it is impossible to $\epsilon$-close reconstruct the characteristic function with probability lower than $1-\delta$.
    The solid lines are log-linear fits of the experimental data. The gray region denotes the sample complexity overhead impossible to achieve without using entanglement.
    (d) Probability of achieving an $\epsilon$-close reconstruction of the $\SI{-4.78}{dB}$, 30-mode characteristic function for various directions in the displacement hyperspace. The shading highlights the proximity to the displacement direction $\beta_0$, points lying on an edge belong to its left region. Each probability is computed using the minimum number of samples required for an $\epsilon$-close reconstruction with a success probability of $1-\delta$ in the $\beta_0$ direction, as shown in (c), with $n=30$ modes and giving $N=1472$ samples. The dashed line indicates the target probability of $1-\delta=\frac{2}{3}$. }
    \label{fig-sm:chn-recon}
\end{figure}

As the first test, we measure $10^6$ samples of the $n=16$-mode displacement process applied to a vacuum input state. Then, for $N=\{2\times10^2, 5\times10^3, 1\times10^5\}$, we randomly select $N$-samples from all the measured samples and run the process reconstruction to estimate $\tilde{\lambda}(\beta)$ on the line segment $\beta=(b, b, \ldots, b), b\in[0, 0.5]$, and repeat the procedure for 25 times (allowing replacement) to determine the estimator's mean value and standard deviation along the ling segment. The effect is shown in \autoref{fig-sm:chn-recon}(a). The most striking observation is that the absolute value of the estimator grows unreasonably at some point, which we colloquially call ``diverge''. The reason is that the estimator has a prefactor of $e^{2ne^{-2r}b^2}$ and amplifies the statistics noises in the Bell measurement for $b$ large. We find that an increasing number of samples improves the fidelity of the reconstructed process and makes the diverge to happen for a larger $b$. However, pushing the diverging point by a fixed amount requires exponentially more samples. This scaling behavior is consistent with theoretical expectations (\autoref{th:est}).

Secondly, we characterize the scaling behavior of squeezing by measuring $25\times10^5$ samples of the $30$-mode displacement process for different squeezing parameters of the two-mode squeezed vacuum state, and run the same resampling procedure as described above to extract the mean value and standard deviation along the same line segment. We compare the reconstructions for three values squeezing: $\SI{0}{dB}$ (vacuum input same as the last test), $\SI{-2.26}{dB}$, and $\SI{-4.78}{dB}$. For each squeezing, we average the reconstruction using $25$ sets of $10^5$ samples and calculate the standard deviation of the reconstruction. These reconstruction results are presented in \autoref{fig-sm:chn-recon}(b). The sample number overhead is greatly relieved by a moderate amount of squeezing. Indeed, whereas the reconstruction of a 30-mode displacement process via vacuum heterodyne detection necessitates a vast number of samples, employing a two-mode squeezed vacuum state enables the reconstruction of the same displacement process with a substantially reduced sample size. The effect is also expected, as a higher squeezing reduces the uncertainty in the measured displacement and increases the information obtained per measurement, leading to improved reconstruction fidelity for larger squeezing values. 

\subsection{Quantitative results}

To investigate the scaling with the number of modes in the process, we measure the two-mode squeezed vacuum state for different $n$-mode displacement processes and compare the results for three distinct squeezing values, $\SI{0}{dB}$, $\SI{-2.26}{dB}$ and $\SI{-4.78}{dB}$. We then introduce the concept of $(\epsilon, \delta)$-close reconstruction to quantify the quality of process reconstruction in a statistical way. 

\begin{definition}[$(\epsilon, \delta)$-close reconstuction]\label{def:eps-close}
    The process reconstruction is $(\epsilon, \delta)$-close to the ground truth $\lambda$ if, after sufficiently many runs of reconstruction using $N$ samples in each run, the estimator $\tilde{\lambda}$ satisfies $|\tilde{\lambda}(\beta)-\lambda(\beta)|<\epsilon$ for any $\beta$ up to a radius $|\beta|<\beta_0$ with a probability of $1-\delta$. 
\end{definition}
We call $N$ the sample complexity of the $(\epsilon, \delta)$-close reconstruction; its scaling behavior with the process mode number will be the primary object of interest in this part of the experiment. In our experiment, we set $\epsilon = 0.24$ and $\delta = \frac{1}{3}$. 

Before proceeding to the technical details of how the sample complexity is obtained, let us briefly comment on the relation between \autoref{def:eps-close} and \autoref{th:main3}. Learning the displacement up to $|\beta|^2\leq\kappa n$ implies being able to $(\epsilon, \delta)$-close reconstruct any process up to the same radius. On the other hand, it is not possible to go the other way and \textit{prove} the learnability by $(\epsilon, \delta)$-close reconstruct many processes, as the number of possible processes is infinite. However, as the lower bound is derived with three-peak process, the sample complexity for experimentally learning a three-peak process and the lower bound can still be comparable. The comparison must be made on the assumption that, the cost of learning the particular three-peak process is the same as that of learning the average-case three-peak processes, whose peak location follows the same distribution. The point will be made clear in the hypothesis testing section.

Another practical challenge for certifying a $(\epsilon, \delta)$-close reconstruction is that, due to the high-dimensionality of the characteristic function space, the computational overhead of a full reconstruction is enormous. For example, for the full reconstruction of a 20-mode process up to $\kappa=0.18$ (radius 0.30) using a mesh size of 0.01, the number of points to be evaluated in the hypersphere is more than $10^{110}$. Fortunately, we have noticed that the process reconstruction of the three-peak process comes with the most error in the direction where the peak exists. In the other directions, the landscape is flat and the divergence happens at larger radius. Therefore, we use the reconstruction in the direction where the peaks exist to calculate the sample complexity, and then randomly sample along all the other directions to show that the same number of samples is sufficient for a better reconstruction in all the other directions.

\begin{algorithm}
    \caption{Determination of the $(\epsilon, \delta)$-close process reconstruction sample complexity}\label{alg:mc-chn-recon}
    \KwData{Bell measurement data ${\bm\zeta} = \{\zeta\}$, $|{\bm \zeta}|=N_{\rm max}$}
    \KwResult{Required sample complexity for $(\epsilon, \delta)$-close process reconstruction up to radius $\beta_0$}
    $N\gets 8e^{2e^{-2r_\textnormal{eff}}|\beta_0|^2}\epsilon^{-2}\log 4\delta^{-1}$\Comment*[r]{Initial guess is the Hoeffding bound}
    $K=25, \quad \mathsf{rec} = \{\}, \quad \mathsf{max\_round} = 35$\Comment*[r]{$K$-repetitions for each process}
    \For{$j\leftarrow 1$ \KwTo $\mathsf{max\_round}$}{
        Sample measurement data ${\bm\zeta}_{(k)}\subseteq{\bm\zeta}, |{\bm\zeta}_{(k)}|=N, k\in\{1,\ldots,K\}$\;
        {$\mathsf{prob\_now} = 0$}\;
        \For{$k\leftarrow 1$ \KwTo $K$}{
        {Estimate $\tilde{\lambda}(\beta), \beta=(b, b, \ldots, b), b\in[0, b_0]$ with ${\bm\zeta}_{(k)}$}\Comment*[r]{Run process recon with Eq.\,\eqref{eq:estimator}}
        $\mathsf{prob\_now} \gets \mathsf{prob\_now}\,+\,1/25$ \textbf{if} $|\tilde{\lambda}(\beta) - \lambda(\beta)| \leq \epsilon,\,\forall \beta$}
        \eIf{$\mathsf{prob\_now} \geq(1-\delta)$}{$N \gets N*e/3$\Comment*[r]{Success prob low - increase number of samples}}{$N \gets N*e/2$\Comment*[r]{Success prob high - decrease number of samples}}
        \lIf{$j>10$}{$\mathsf{rec}$.append(N)\Comment*[r]{Note down those sample complexity close to true value}}
    }
    \Return $\mathsf{rec}.{\rm mean}()$, $\mathsf{rec}.{\rm std}()$
\end{algorithm}

We calculate the sample cost using a Monte Carlo routine, described in \autoref{alg:mc-chn-recon}. The idea of the algorithm is to (1) resample from a pool of measured data for several times, (2) run process reconstruction with Eq.\,\eqref{eq:estimator} for each of the samples, (3) calculate the success probability of the $(\epsilon, \delta)$-close reconstruction along the line segment ${\beta}_0 = b\,(1, 1, \ldots, 1),\; b\in[0, 0.3]$, and (4) dynamically adjust the resampling size until the success probability converges to the target value.
At the end of the algorithm, we estimate the required sample complexity as the average sample number in the last 25 rounds of adjustments, 
\begin{align}
    \bar{N} = \sum^{25}_i N_i.
\end{align}
To determine the uncertainty of the sample complexity, we correct the standard deviation of the sample number in the last 25 rounds of adjustments by a factor $\sqrt{\dfrac{N_\text{max}}{N_\text{max}-\bar{N}}}$:
\begin{equation}
    \Delta N = \sqrt{\frac{N_\text{max}}{N_\text{max}-\bar{N}}}\sqrt{\frac{\sum^{25}_i(\bar{N}-N_i)^2}{25}}.
\end{equation}
Here, $N_{\rm max}$ is the pool size or the total number of measured data. The factor thus takes into account the finite sample size in the bootstrapping process.

In Fig.\,\ref{fig-sm:chn-recon}(c), we compare the reconstruction sample complexity scaling against the number of modes with the cost lower bound defined in \autoref{th:main3}. For both the entanglement-assisted and the conventional case, the cost scales exponentially with the number of modes involved in the displacement process. However, increasing the squeezing allows for a slower scaling in the cost and proves a large polynomial quantum advantage. For a squeezing of $\SI{-4.78}{dB}$ the cost needed to reconstruct allows us to beat the lower bound from a $30$-mode displacement process. For $100$ modes, the lower bound cost is $2.8\times10^{12}$ whereas the cost for the largest squeezing is approximately $10^7$. This leads to an improvement of $10^5$. Considering that the bound is not tight, reconstruction with a squeezing of $\SI{-4.78}{dB}$ gives an empirical improvement of $10^{11.8}$ compared to the vacuum heterodyne method. We also compare the measured sample complexity with the Hoeffding bound (Eq.\,\eqref{eq:upper_bound}). All the data point lies reasonably under the Hoeffding bound, with the fitting curve slopes similar as the bound predicts.

Further, in Fig.\,\ref{fig-sm:chn-recon}(d) we verify that the reconstruction is hardest along those directions $\beta$ where $\langle\beta_0|\beta\rangle$ is large. This verification is necessary because it provides evidence that the estimated sample complexity on the direction ${\beta}_0 = b\,(1, 1, \ldots, 1)$ is the highest across the entire hyperball. We implement this verification by running the same process reconstruction algorithm (\autoref{alg:mc-chn-recon}), but along another direction $\beta$ in the dual space. We run the test both for the random choice of directions, which in the high mode number limit results in $\braket{\beta_0|\beta}\to0$, and for $7$ other $\beta$ close to the $\beta_0$-direction: $\beta=\{0.40, 0.50, 0.60,0.70,0.80,0.90, 0.95\}$. We found that the randomly chosen $\beta$ directions results in a success probability of $0.77\pm 0.04$. In comparison, as $\braket{\beta_0|\beta}$ increases the probability starts to decline. None of the numerical probabilities fall below the threshold $1-\delta=0.67$ by more than one standard deviation. The behavior is as expected, and supports the argument that the estimated quantum improvement from the ratios of sample complexity is conservative. 

\section{Hypothesis testing}

Although we observed an improvement in the process reconstruction task due to the adoption of quantum entanglement, we cannot claim this improvement as a provable quantum advantage. \autoref{th:main2} establishes a lower bound on the sample complexity of a scheme that can learn the characteristic function $\lambda(\beta)$ accurately for any random displacement process in a large family, and for all values of $\beta$ in a specified bounded range. But our experiment learns the characteristic function for processes chosen from a smaller family and for more restricted values of $\beta$.

To establish a provable quantum advantage, we experimentally realize the hypothesis-testing game described in the proof of \autoref{th:main3}. This procedure is feasible because it suffices to reconstruct processes in the three-peak family $\Lambda^{\epsilon,\sigma}_\text{3-peak}$. Winning the hypothesis-testing with significant success probability by conducting a number of entanglement-enhanced experiments far less than the entanglement-free lower bound rigorously establishes a large quantum advantage.


\subsection{Protocol}

\begin{figure*}[ht]
    \centering
    \includegraphics[width=\textwidth]{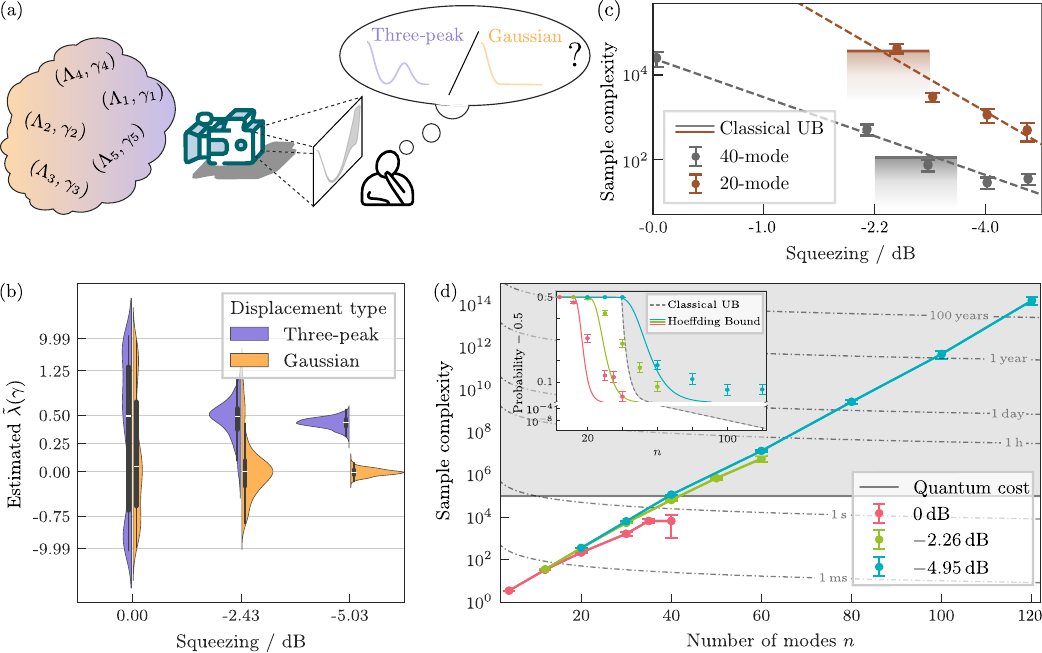}
    \caption{\textbf{Hypothesis testing.} (a) The objective is to distinguish whether a displacement process belongs to the three-peak family with an unknown parameter or the Gaussian family. (b) An example of the separation of the estimator, $\tilde{\lambda}(\beta=\gamma)$, for two types of 40-mode displacement processes using different amounts of squeezing. (c) Sample complexity required to achieve $2/3$ success probability in a $\kappa=0.2$ hypothesis test, measured with varying amounts of squeezing. The classical complexity bound (solid) for achieving the same success probability and the exponential fit (dashed) are also shown. (d) Inset: the measured probability of winning the $\kappa=0.2$ hypothesis testing game for different numbers of modes, using $10^5$ samples and various amounts of squeezing. Solid lines represent a pessimistic estimation of success probability derived from the Hoeffding bound\,\cite{Oh2024}.
    Main: Minimum sample complexity for any conventional strategy to achieve the same success probability as reported in the inset, calculated according to the classical complexity bound, and the corresponding sample collection time at a 1\,MHz bandwidth.
    Error bars represent the $1\sigma$ standard deviation from a 25-step sequential sampling. 
    The shaded region indicates the existence of a quantum advantage.}
    \label{fig-sm:hypo-test}
\end{figure*}

A hypothesis test distinguishes an $(1,n)$-mode Gaussian displacement process $\Lambda_0$ from an $(1,n)$-mode three-peak displacement process $\Lambda_\gamma$, where the displacement direction $\gamma$ in the dual space is random and follows a specific distribution. The game, as illustrated in \autoref{fig-sm:hypo-test}, involves a dealer and a challenger and proceeds as follows:

\begin{enumerate}
    \item The dealer randomly selects the types of $K$ displacement processes $\Lambda_k, k\in\{1, \ldots, K\}$. The types can be three-peak processes $\Lambda_\gamma$ (\autoref{def:3-peak-chn}) or Gaussian processes $\Lambda_0$ (\autoref{def:gaussian-chn}).

    \item For every three-peak process, the dealer draws a vector $\gamma \in \mathbb{C}^n$ according to the multivariate normal distribution $q(\gamma)$:
    \begin{equation}
        q(\gamma)=\left(\frac{1}{2\pi\sigma_\gamma^2}\right)^ne^{-\frac{|\gamma|^2}{2\sigma_\gamma^2}}.
    \end{equation}

    \item The Gaussian process does not depend on $\sigma_\gamma$ and shares the same parameter $\sigma$ as the three-peak processes. However, a fictional $\gamma$ is drawn from the same distribution and assigned to the Gaussian process.

    \item The dealer prepares the processes according to the parameters decided above. They then allow a challenger to use each process $N$ times without revealing the type or the $\gamma$ value's for the three-peak processes. The challenger uses their preferred strategy to measure the output of the process.

    \item After all the measurements are completed, the dealer announces $\gamma_k,\;k\in\{1, \ldots, K\}$. The challenger is then asked to identify the class of the process using their measurement outcomes.
\end{enumerate}
In our experiment, we always choose the parameters as $K=16$, $\sigma=0.3$, $\kappa=0.2$, and $2\sigma_\gamma^2=0.99\kappa$. The characteristic function of the three-peak process has two extra peaks at locations $\pm\gamma_k$; the height of the side peaks, appearing in \autoref{def:3-peak-chn}, is $2\epsilon_0=0.5$. This feature is absent for the Gaussian process. The challenger learns the type of the process by identifying whether the two extra peaks are present or absent.

To determine the type of each process, a quantum challenger with access to entangled resources will prepare $n$-pairs of two-mode squeezed state (EPR pairs) for each implementation of the process. They take one mode from each EPR pair to form an $n$-mode probe state and apply the process to it. They then perform Bell measurement on each (now displaced) EPR pair at the end of the process and record the outcomes. Once the value of $\gamma$ is announced, they calculate the estimator $\tilde{\lambda}(\beta)$ at $\beta=\gamma$ and compare it with the threshold $\lambda_0=0.25$ to classify the process. If $\tilde{\lambda}(\beta)>\lambda_0$, they will classify the process as three-peak; if $\tilde{\lambda}(\beta)\leq\lambda_0$, they will classify the process as Gaussian. For a conventional challenger without access to entanglement, the probe state reduces to a vacuum state, and the Bell measurement becomes a heterodyne measurement.

Before proceeding further, let us remark on our parameter choices in the game. For certain values of $|\gamma|$, no challenger can distinguish the type of the processes. Specifically, if $|\gamma|^2<2\sigma^2$, the side peaks in the three-peak distribution will be buried in the central peak, making the total variation distance between the two types of processes too small. In this case, no strategy will be able to distinguish the type. 
Due to the heavy concentration of the multi-mode $\chi^2$-distribution toward the expectation value, we did not encounter this exception case in the experiment for a single time.
Secondly, for very large $|\gamma|^2>\kappa n $, the estimator is expected to be dominated by statistical noise, exhibiting ``divergence'' behavior observed in the process reconstruction test. In this scenario, learning becomes equivalent to random guessing. By specifying $2\sigma_\gamma^2=0.99\kappa$\,\cite{Oh2024}, we limit the probability of this occurrence to below $50\%$. This ensures that a fair number of displacement processes will be of the non-diverging and learnable type, allowing the game to reveal the performance differences between challengers with and without access to quantum entanglement.

\subsection{Results}
We again start by visualizing how the hypothesis testing will be able to demonstrate quantum advantage. To this end, we use different two-mode squeezed vacuum states with squeezing levels of $\SI{0}{dB}$, $\SI{-2.26}{dB}$ and $\SI{-4.95}{dB}$ as input states. We conduct hypothesis testing on a set of $n=30$-mode processes (8 three-peak, 8 Gaussian) using $10^5$ samples. For each process, we repeat the protocol 25 times, calculate the estimators according to  Eq.\,\eqref{eq:estimator}, and display the clustering behavior of the estimators in a violin plot as shown in \autoref{fig-sm:hypo-test}(b). 
We see that the center of the distribution for all obtained estimators is close to the ground truth values ($\approx0.5$ for three-peak and 0 for Gaussian processes). However, without squeezing, the large variance causes the two distributions to overlap, making it impossible to distinguish between them using the estimator. As squeezing increases, the variance rapidly decreases, and the distributions converge to the expectation values. 
At a squeezing level of $\SI{-4.95}{dB}$, the support of the two distributions becomes completely disjoint, allowing the process types to be learned with certainty. Note that the deviation of the mean value of the $\SI{-4.95}{dB}$-squeezed data from the ground truth is due to the slightly increased phase noise in the specific experiment. It causes the peak to shift from the ideal location by a bit. This way, the estimator $|\lambda(\beta = \gamma)|$ did not ``catch'' the top of the peak, causing all the numbers to be smaller than in theory. On the other hand, the estimator itself remains unbiased for any squeezing level.

\begin{algorithm}
    \caption{Determination of the hypothesis testing sample complexity}\label{alg:mc-hypo-test}
    \KwData{Bell measurement data $\mathcal{Z} = \{\bm\zeta\}$, ${\bm\zeta}^{[m]} = \{\zeta\}^{[m]}$, $\min_{m=1}^M |{\bm \zeta^{[m]}}|=N_{\rm max}$ \\
    \hspace{33pt} Process type ground truth $\{T_m\}$, $T_m=1$(0) $\Leftrightarrow$ Process $m$ is three-peak (Gaussian) \\
    \hspace{33pt} Displacement specifications $\{\gamma_m\}, m\in\{1,\ldots,M\}$. \\ \hspace{33pt} Number of processes $M=16$. Target success probability is $P_{\rm suc}$.}
    \KwResult{Required sample complexity for hypothesis testing to achieve target success probability}
    $N\gets 8e^{2e^{-2r_\textnormal{eff}}|\beta_0|^2}\epsilon_0^{-2}\log 4(1-P_{\rm suc})^{-1}, \quad \epsilon_0=0.25$\Comment*[r]{Initial guess is the Hoeffding bound}
    $K=25, \quad \mathsf{rec} = \{\}, \quad \mathsf{max\_round} = 35$\Comment*[r]{$K$-repetitions for each process}
    \For{$j\leftarrow 1$ \KwTo $\mathsf{max\_round}$}{
        {$\mathsf{prob\_now} = 0$}\;
        \For{$m\leftarrow1$ \KwTo $M$}{
            Sample measurement data ${\bm\zeta}_{(k)}^{[m]}\subseteq{\bm\zeta}, \left|{\bm\zeta}_{(k)}^{[m]}\right|=N, k\in\{1,\ldots,K\}$\;
            \For{$k\leftarrow 1$ \KwTo $K$}{
            {Calculate $\tilde{\lambda}(\beta),\;\beta=\gamma$ with ${\bm\zeta}_{(k)}^{[m]}$}\Comment*[r]{Calculate estimator value with Eq.\,\eqref{eq:estimator}}
            $\mathsf{prob\_now} \gets \mathsf{prob\_now}\,+\,1/(MK)$ \textbf{if} $(\tilde{\lambda}(\beta)>\epsilon\;\mathsf{and}\;T_m=1)\;\mathsf{OR}\;(\tilde{\lambda}(\beta)<\epsilon\;\mathsf{and}\;T_m=0)$
            \Comment*[r]{Guessing is correct if $\tilde{\lambda}(\beta)>(<)0.25$ for three-peak (Gaussian) process}}}
        \eIf{$\mathsf{prob\_now} \geq P_{\rm suc}$}{$N \gets N*e/3$\Comment*[r]{Success probability low - increase number of samples}}{$N \gets N*e/2$\Comment*[r]{Success probability high - decrease number of samples}}
        \lIf{$k>10$}{$\mathsf{rec}$.append(N)\Comment*[r]{Note down those sample complexities close to true value}}
    }
    \Return $\mathsf{rec}.{\rm mean}()$, $\mathsf{rec}.{\rm std}()$
\end{algorithm}

Secondly, we investigate the relation between the sample complexity required to achieve a specific success probability of learning and the input squeezing level. We study this relation with a family of 20-mode processes and a family of 40-mode processes, setting the target success probability to $P_{\rm suc} = 2/3$. To determine the number of samples required, we again use a Monte Carlo routine as described in \autoref{alg:mc-hypo-test}. This idea is akin to the algorithm for determining the sample complexity for $(\epsilon, \delta)$-close process reconstruction, except that the estimator is calculated for only one point $\beta=\gamma$, and the success probability is averaged over all 16 processes.

For the 20-mode experiment, we set the input squeezing levels to $\SI{0}{dB}$ (vacuum), $\SI{2.12}{dB}$, $\SI{2.97}{dB}$, $\SI{4.01}{dB}$ and $\SI{4.90}{dB}$, and determine the sample complexity. For the 40-mode experiment, the squeezing levels are $\SI{2.52}{dB}$, $\SI{3.05}{dB}$, $\SI{4.01}{dB}$ and $\SI{4.87}{dB}$. It is not possible to run the 40-mode hypothesis testing with the conventional strategy. We carefully adjusted the squeezing levels to align with integer values. However, the fluctuation of the setup over time prevents us from maintaining consistent squeezing throughout a long (40-minute) measurement. We report the results of the sample complexity scaling with squeezing in \autoref{fig-sm:hypo-test}(b). The data points generally conform well with the predicted scaling behavior in Eq.\,\eqref{eq:upper_bound}, indicating that the required number of samples is inversely proportional to the exponential of the fluctuation of the input state, measured by its variance. Therefore, squeezing dramatically decreases the sample complexity overhead for hypothesis testing.

Interestingly, a provable quantum advantage can already be found in these intermediate-scale hypothesis testing tasks. Using Eq.\,\eqref{eq:prob_tvd}, we can calculate the number of samples that a conventional scheme would need to achieve the same success probability of learning, which turns out to be:
\begin{align}
    N_{\rm c} =\frac{2P_{\rm suc}-1}{16\,\epsilon_0^2} \left(1+\frac{1.98\kappa}{1+2\sigma^2}\right)^n.
    \label{eq:n_crit}
\end{align}
Substituting the experimental parameters into Eq.\,\eqref{eq:ub_tvd}, we determine this critical complexity to be $1.20\times10^2$ ($3.93\times10^4$) for the 20(40)-mode learning task. In contrast, the lowest sample complexity we measured using squeezing is 30.4 and 507 respectively. Thus, we find quantum advantages with factors of 3.9 and 77.4. Indeed, the quantum advantage increases for the learning of a man-mode process.

Thirdly, we characterize the scaling of quantum advantage against the number of modes for a fixed-size hypothesis testing game. A possible way of achieving this is to measure the challenger's success probability and directly compare it with the best that a conventional strategy can achieve with the same number of samples, $N=10^5$. By inserting Eq.\,\eqref{eq:ub_tvd} into Eq.\,\eqref{eq:prob_tvd}, we find the success probability of a conventional strategy is upper bounded by:
\begin{align}
    P_{\rm c} =\frac{1}{2} \left(1+16N\epsilon_0^2\left(\frac{1+2\sigma^2}{1+2\sigma^2+1.98\kappa}\right)^n\right).
    \label{eq:p_crit}
\end{align}
With these measured success probabilities, an equivalent but more intuitive way of stating the results is to convert these probabilities into the equivalent classical complexity required for a conventional learning strategy---exactly as done in Eq.\,\eqref{eq:n_crit}, and compare the classical complexity with the number of samples actually used by a quantum learner. This way, the ratio of the two sample complexity quantifies the amount of quantum advantage. 

We have conducted several hypothesis testing with different squeezing and number of modes. Each hypothesis testing consisting uses a resolution constant $\kappa=0.2$ and is composed of 16 processes. For each process, we send $N_{\rm max}=10^6$ probe states through the displacement process, collect the Bell measurement results and obtain a pool of the estimator $\lambda(\beta=\gamma)$ values. We then obtain $R=25$ repeated samplings of the estimators, calculate their average value, $\bar\lambda$, and the standard deviation, $\overline{\lambda^2}-\bar\lambda^2$. Using the central limit theorem, the probability of correctly guessing the process type as three-peak (Gaussian) is:
\begin{align}
    \bar{P} = \frac{1}{2}\pm\frac{1}{2}{\rm erf}\left(\frac{\bar\lambda-\epsilon_0}{\sqrt{2(\overline{\lambda^2}-\bar\lambda^2)}}\right).
\end{align}
As the guessing results of the process type is a binary event, we calculate the uncertainty on the probability using the standard deviation of a binomial distribution multiplied by a factor to correct the effect from bootstrapping: 
\begin{equation}
    \Delta P=\sqrt{\frac{\bar{P}(1-\bar{P})}{R}}\sqrt{\frac{N}{N_\text{max}-N}}.
\end{equation}
We obtain the error bar of each data point via error propagation of the results measured for all the processes used in the hypothesis testing.

In \autoref{fig-sm:hypo-test}(d) inset, we present the results of hypothesis testing raw success probability with the number of modes. We found that increasing the mode number makes the guessing increasingly challenging, as the ``divergence'' in the process reconstruction appears faster for smaller values of $|\beta|$. Regardless of the input state, the success probabilities for the learning of few-mode processes are close to $100 \%$ as the noise is almost negligible for $\beta = \gamma$. 
Above a specific mode number, the success probability rapidly decreases to $50\%$ as the estimator diverges to infinity at $\beta = \gamma$. The transition point depends on the squeezing level of the two-mode squeezed state: a higher squeezing causes the transition to occur at a higher number of modes. Specifically, for squeezing levels of $\SI{0}{dB}$, $\SI{-2.26}{dB}$ and $\SI{-4.95}{dB}$, the ``critical'' mode numbers where the guessing becomes imperfect are $16$, $20$ and $40$ modes, respectively. Greater entanglement at higher squeezing levels improves the success rate in discriminating the process type. 
Moreover, even only a small amount of squeezing ($\SI{-2.26}{dB}$) is used, the success probability of hypothesis testing can exceed the classical upper bound, Eq.\,\eqref{eq:p_crit}, at several data points for high mode number processes $d>40$. This indicates the existence of a provable quantum advantage. For the experiment with higher input squeezing ($\SI{-4.95}{dB}$), the success probability is even higher, and the quantum advantage remains statistically significant ($>99.3\%$ confidence) from $n=60$-mode all the way to $n=120$. On the other hand, as expected, the conventional learning strategy is indeed significantly lower than the theoretical upper bound. This emphasizes the power of entanglement for quantum-enhanced learning over any conventional method.


Finally, we convert the success probability to the equivalent classical complexity to quantify the quantum advantage. In doing this, we map the average and the $\pm 1\sigma$ probability in the inset to the sample complexity using Eq.\,\eqref{eq:n_crit}. The results are shown in \autoref{fig-sm:hypo-test}(d). Note that if the success probability is not significant, the equivalent sample complexity will become negative, resulting in an infinitely long error bar (shown as a caret arrow) in the log-scale plot. 
Due to the scaling behavior of the sample complexity, we are able to certify a large amount of quantum advantage in high mode number experiments.
For the largest scale (120-mode) experiment, we measured a hypothesis testing success probability of $0.563\pm 0.025$, exceeding the bound for conventional strategy ($0.5+3.8\times10^{-11}$) with a confidence level of $99.3\%$. To learn one process and achieve the same success probability with a conventional approach, $1.6\times10^{14}$ classical samples would be required. This translates into an expected increase in measurement time of more than 600 years if using a bandwidth of 1\,MHz---same as in our experiment. Our result demonstrates a quantum advantage of $9.2$ orders of magnitude.

\end{document}